\documentclass[12pt]{article}
\usepackage{graphicx}
\usepackage{epstopdf}
\usepackage{subfig}
\usepackage{caption}
\usepackage{mathrsfs}
\usepackage{amssymb}
\usepackage{amsmath}
\usepackage{verbatim}
\usepackage{epsfig}
\usepackage[bookmarks]{hyperref}
\advance\textheight by \topskip
\begin{document}


\newcommand{\HPA}[1]{{\it Helv.\ Phys.\ Acta.\ }{\bf #1}}
\newcommand{\AP}[1]{{\it Ann.\ Phys.\ }{\bf #1}}
\newcommand{\be}{\begin{equation}}
\newcommand{\ee}{\end{equation}}
\newcommand{\br}{\begin{eqnarray}}
\newcommand{\er}{\end{eqnarray}}
\newcommand{\ba}{\begin{array}}
\newcommand{\ea}{\end{array}}
\newcommand{\bi}{\begin{itemize}}
\newcommand{\ei}{\end{itemize}}
\newcommand{\bn}{\begin{enumerate}}
\newcommand{\en}{\end{enumerate}}
\newcommand{\bc}{\begin{center}}
\newcommand{\ec}{\end{center}}
\newcommand{\ul}{\underline}
\newcommand{\ol}{\overline}
\def\l{\left\langle}
\def\r{\right\rangle}
\def\as{\alpha_{s}}
\def\ycut{y_{\mbox{\tiny cut}}}
\def\yij{y_{ij}}
\def\epem{\ifmmode{e^+ e^-} \else{$e^+ e^-$} \fi}
\newcommand{\eeww}{$e^+e^-\rightarrow W^+ W^-$}
\newcommand{\qqQQ}{$q_1\bar q_2 Q_3\bar Q_4$}
\newcommand{\eeqqQQ}{$e^+e^-\rightarrow q_1\bar q_2 Q_3\bar Q_4$}
\newcommand{\eewwqqqq}{$e^+e^-\rightarrow W^+ W^-\ar q\bar q Q\bar Q$}
\newcommand{\eeqqgg}{$e^+e^-\rightarrow q\bar q gg$}
\newcommand{\eeqloop}{$e^+e^-\rightarrow q\bar q gg$ via loop of quarks}
\newcommand{\eeqqqq}{$e^+e^-\rightarrow q\bar q Q\bar Q$}
\newcommand{\eewwjjjj}{$e^+e^-\rightarrow W^+ W^-\rightarrow 4~{\rm{jet}}$}
\newcommand{\eeqqggjjjj}{$e^+e^-\rightarrow q\bar
q gg\rightarrow 4~{\rm{jet}}$}
\newcommand{\eeqloopjjjj}{$e^+e^-\rightarrow q\bar
q gg\rightarrow 4~{\rm{jet}}$ via loop of quarks}
\newcommand{\eeqqqqjjjj}{$e^+e^-\rightarrow q\bar q Q\bar Q\rightarrow
4~{\rm{jet}}$}
\newcommand{\eejjjj}{$e^+e^-\rightarrow 4~{\rm{jet}}$}
\newcommand{\jjjj}{$4~{\rm{jet}}$}
\newcommand{\qqbar}{$q\bar q$}
\newcommand{\ww}{$W^+W^-$}
\newcommand{\ar}{\rightarrow}
\newcommand{\sm}{${\cal {SM}}$}
\newcommand{\Dir}{\kern -6.4pt\Big{/}}
\newcommand{\Dirin}{\kern -10.4pt\Big{/}\kern 4.4pt}
\newcommand{\DDir}{\kern -8.0pt\Big{/}}
\newcommand{\DGir}{\kern -6.0pt\Big{/}}
\newcommand{\wwqqqq}{$W^+ W^-\ar q\bar q Q\bar Q$}
\newcommand{\qqgg}{$q\bar q gg$}
\newcommand{\qloop}{$q\bar q gg$ via loop of quarks}
\newcommand{\qqqq}{$q\bar q Q\bar Q$}

\def\st{\sigma_{\mbox{\scriptsize t}}}
\def\Ord{\buildrel{\scriptscriptstyle <}\over{\scriptscriptstyle\sim}}
\def\OOrd{\buildrel{\scriptscriptstyle >}\over{\scriptscriptstyle\sim}}
\def\jhep #1 #2 #3 {{JHEP} {\bf#1} (#2) #3}
\def\plb #1 #2 #3 {{Phys.~Lett.} {\bf B#1} (#2) #3}
\def\npb #1 #2 #3 {{Nucl.~Phys.} {\bf B#1} (#2) #3}
\def\epjc #1 #2 #3 {{Eur.~Phys.~J.} {\bf C#1} (#2) #3}
\def\zpc #1 #2 #3 {{Z.~Phys.} {\bf C#1} (#2) #3}
\def\jpg #1 #2 #3 {{J.~Phys.} {\bf G#1} (#2) #3}
\def\prd #1 #2 #3 {{Phys.~Rev.} {\bf D#1} (#2) #3}
\def\prep #1 #2 #3 {{Phys.~Rep.} {\bf#1} (#2) #3}
\def\prl #1 #2 #3 {{Phys.~Rev.~Lett.} {\bf#1} (#2) #3}
\def\mpl #1 #2 #3 {{Mod.~Phys.~Lett.} {\bf#1} (#2) #3}
\def\rmp #1 #2 #3 {{Rev. Mod. Phys.} {\bf#1} (#2) #3}
\def\cpc #1 #2 #3 {{Comp. Phys. Commun.} {\bf#1} (#2) #3}
\def\sjnp #1 #2 #3 {{Sov. J. Nucl. Phys.} {\bf#1} (#2) #3}
\def\xx #1 #2 #3 {{\bf#1}, (#2) #3}
\def\hepph #1 {{\tt hep-ph/#1}}
\def\preprint{{preprint}}

\def\beq{\begin{equation}}
\def\beeq{\begin{eqnarray}}
\def\eeq{\end{equation}}
\def\eeeq{\end{eqnarray}}
\def\a0{\bar\alpha_0}
\def\thrust{\mbox{T}}
\def\Thrust{\mathrm{\tiny T}}
\def\ae{\alpha_{\mbox{\scriptsize eff}}}
\def\ap{\bar\alpha_p}
\def\as{\alpha_{\mathrm{S}}}
\def\aem{\alpha_{\mathrm{EM}}}
\def\b0{\beta_0}
\def\cN{{\cal N}}
\def\cd{\chi^2/\mbox{d.o.f.}}
\def\Ecm{E_{\mbox{\scriptsize cm}}}
\def\ee{e^+e^-}
\def\enap{\mbox{e}}
\def\eps{\epsilon}
\def\ex{{\mbox{\scriptsize exp}}}
\def\GeV{\mbox{\rm{GeV}}}
\def\half{{\textstyle {1\over2}}}
\def\jet{{\mbox{\scriptsize jet}}}
\def\kij{k^2_{\bot ij}}
\def\kp{k_\perp}
\def\kps{k_\perp^2}
\def\kt{k_\bot}
\def\lms{\Lambda^{(n_{\rm f}=4)}_{\overline{\mathrm{MS}}}}
\def\mI{\mu_{\mathrm{I}}}
\def\mR{\mu_{\mathrm{R}}}
\def\MSbar{\overline{\mathrm{MS}}}
\def\mx{{\mbox{\scriptsize max}}}
\def\NP{{\mathrm{NP}}}
\def\pd{\partial}
\def\pt{{\mbox{\scriptsize pert}}}
\def\pw{{\mbox{\scriptsize pow}}}
\def\so{{\mbox{\scriptsize soft}}}
\def\st{\sigma_{\mbox{\scriptsize tot}}}
\def\ycut{y_{\mathrm{cut}}}
\def\slashchar#1{\setbox0=\hbox{$#1$}           
     \dimen0=\wd0                                 
     \setbox1=\hbox{/} \dimen1=\wd1               
     \ifdim\dimen0>\dimen1                        
        \rlap{\hbox to \dimen0{\hfil/\hfil}}      
        #1                                        
     \else                                        
        \rlap{\hbox to \dimen1{\hfil$#1$\hfil}}   
        /                                         
     \fi}                                         %
\def\etmiss{\slashchar{E}^T}
\def\Meff{M_{\rm eff}}
\def\Ord{\lsim}
\def\OOrd{\gsim}
\def\tq{\tilde q}
\def\tchi{\tilde\chi}
\def\lsp{\tilde\chi_1^0}

\def\gam{\gamma}
\def\ph{\gamma}
\def\be{\begin{equation}}
\def\ee{\end{equation}}
\def\bea{\begin{eqnarray}}
\def\eea{\end{eqnarray}}
\def\lsim{\:\raisebox{-0.5ex}{$\stackrel{\textstyle<}{\sim}$}\:}
\def\gsim{\:\raisebox{-0.5ex}{$\stackrel{\textstyle>}{\sim}$}\:}

\def\ino{\mathaccent"7E} \def\gluino{\ino{g}} \def\mgluino{m_{\gluino}}
\def\sqk{\ino{q}} \def\sup{\ino{u}} \def\sdn{\ino{d}}
\def\chargino{\ino{\omega}} \def\neutralino{\ino{\chi}}
\def\cab{\ensuremath{C_{\alpha\beta}}} \def\proj{\ensuremath{\mathcal P}}
\def\dab{\delta_{\alpha\beta}}
\def\zz{s-M_Z^2+iM_Z\Gamma_Z} \def\zw{s-M_W^2+iM_W\Gamma_W}
\def\prop{\ensuremath{\mathcal G}} \def\ckm{\ensuremath{V_{\rm CKM}^2}}
\def\aem{\alpha_{\rm EM}} \def\stw{s_{2W}} \def\sttw{s_{2W}^2}
\def\nc{N_C} \def\cf{C_F} \def\ca{C_A}
\def\qcd{\textsc{Qcd}} \def\susy{supersymmetric} \def\mssm{\textsc{Mssm}}
\def\slash{/\kern -5pt} \def\stick{\rule[-0.2cm]{0cm}{0.6cm}}
\def\h{\hspace*{-0.3cm}}

\def\ims #1 {\ensuremath{M^2_{[#1]}}}
\def\tw{\tilde \chi^\pm}
\def\tz{\tilde \chi^0}
\def\tf{\tilde f}
\def\tl{\tilde l}
\def\ppb{p\bar{p}}
\def\gl{\tilde{g}}
\def\sq{\tilde{q}}
\def\sqb{{\tilde{q}}^*}
\def\qb{\bar{q}}
\def\sqL{\tilde{q}_{_L}}
\def\sqR{\tilde{q}_{_R}}
\def\ms{m_{\tilde q}}
\def\mg{m_{\tilde g}}
\def\Gs{\Gamma_{\tilde q}}
\def\Gg{\Gamma_{\tilde g}}
\def\md{m_{-}}
\def\eps{\varepsilon}
\def\Ce{C_\eps}
\def\dnq{\frac{d^nq}{(2\pi)^n}}
\def\DR{$\overline{DR}$\,\,}
\def\MS{$\overline{MS}$\,\,}
\def\DRm{\overline{DR}}
\def\MSm{\overline{MS}}
\def\ghat{\hat{g}_s}
\def\shat{\hat{s}}
\def\sihat{\hat{\sigma}}
\def\Li{\text{Li}_2}
\def\bs{\beta_{\sq}}
\def\xs{x_{\sq}}
\def\xsa{x_{1\sq}}
\def\xsb{x_{2\sq}}
\def\bg{\beta_{\gl}}
\def\xg{x_{\gl}}
\def\xga{x_{1\gl}}
\def\xgb{x_{2\gl}}
\def\lsp{\tilde{\chi}_1^0}

\def\gluino{\mathaccent"7E g}
\def\mgluino{m_{\gluino}}
\def\squark{\mathaccent"7E q}
\def\msquark{m_{\mathaccent"7E q}}
\def\M{ \overline{|\mathcal{M}|^2} }
\def\utm{ut-M_a^2M_b^2}
\def\MiLR{M_{i_{L,R}}}
\def\MiRL{M_{i_{R,L}}}
\def\MjLR{M_{j_{L,R}}}
\def\MjRL{M_{j_{R,L}}}
\def\tiLR{t_{i_{L,R}}}
\def\tiRL{t_{i_{R,L}}}
\def\tjLR{t_{j_{L,R}}}
\def\tjRL{t_{j_{R,L}}}
\def\tg{t_{\gluino}}
\def\uiLR{u_{i_{L,R}}}
\def\uiRL{u_{i_{R,L}}}
\def\ujLR{u_{j_{L,R}}}
\def\ujRL{u_{j_{R,L}}}
\def\ug{u_{\gluino}}
\def\utot{u \leftrightarrow t}
\def\ar{\to}
\def\sqk{\mathaccent"7E q}
\def\sup{\mathaccent"7E u}
\def\sdn{\mathaccent"7E d}
\def\chargino{\mathaccent"7E \chi}
\def\neutralino{\mathaccent"7E \chi}
\def\slepton{\mathaccent"7E l}
\def\M{ \overline{|\mathcal{M}|^2} }
\def\cab{\ensuremath{C_{\alpha\beta}}}
\def\ckm{\ensuremath{V_{\rm CKM}^2}}
\def\zz{s-M_Z^2+iM_Z\Gamma_Z}
\def\zw{s-M_W^2+iM_W\Gamma_W}
\def\s22w{s_{2W}^2}

\newcommand{\cpmtwo}    {\mbox{$ {\chi}^{\pm}_{2}                    $}}
\newcommand{\cpmone}    {\mbox{$ {\chi}^{\pm}_{1}                    $}}

\begin{flushright}
{SHEP-09-29}\\
{DFTT 70/2009}\\
\today
\end{flushright}
\vskip0.1cm\noindent
\begin{center}
{{\large {\bf Strong and Electro-Weak Supersymmetric Corrections \\[0.25cm]
              to Single Top Processes at the Large Hadron Collider}}
\\[1.0cm]
{\large G. Macorini$^1$, S. Moretti$^{2,3}$ and L. Panizzi$^4$}\\[0.30 cm]
{\it  $^1$Dipartimento di Fisica, Universit\`a del Salento and INFN,}\\
{\it  Via Arnesano, 73100 Lecce, Italy.}\\[0.20 cm]
{\it  $^2$School of Physics and Astronomy, University of Southampton,}\\
{\it  Highfield, Southampton SO17 1BJ, UK.}\\[0.20 cm]
{\it  $^3$Dipartimento di Fisica Teorica, Universit\`a di Torino,}\\
{\it  Via Pietro Giuria 1, 10125 Torino, Italy.}\\[0.20 cm]
{\it  $^4$Universit\'e de Lyon, France; Universit\'e Lyon 1, CNRS/IN2P3,}\\
{\it  UMR5822 IPNL, F-69622 Villeurbanne Cedex, France.}
}
\\[1.25cm]
\end{center}

\begin{abstract}
{\small
\noindent
We present the one-loop corrections originating from Quantum Chromo-Dynamics (QCD) and 
Electro-Weak (EW) interactions of Supersymmetric (SUSY) origin within the Minimal Supersymmetric
Standard Model (MSSM) to the single-top processes $bq\to tq'$ and $q\bar q'\to t\bar b$.
We illustrate their impact onto top quark observables accessible at the
Large Hadron Collider (LHC) in the `$t+$jet' final state, such as total cross section,
several differential distributions and 
left-right plus forward-backward asymmetries. We find that in many instances these effects 
can be observable for planned LHC energies and luminosities, quite large as well as rather 
sensitive to several MSSM parameters.
}

\end{abstract}

\newpage


\section{Introduction}
\label{Sec:Intro}
Top quark processes at the LHC will be an ideal laboratory where it will be possible to profile with
great accuracy the heaviest particle of the Standard Model (SM) and eventually extract possible
effects of New Physics (NP) Beyond the SM (BSM). The manifestation of NP at the LHC will in fact
depend on the actual NP mass scale, $M_{\rm{NP}}$. If the available energy is less than
$M_{\rm{NP}}$, then NP will manifests itself through {\it virtual}\, effects. Conversely, if the
available energy is greater than $M_{\rm{NP}}$, NP will appear through the production and decay of
new {\it real} particle resonances.  A mixture of both scenarios may of course also occur.

If NP is identified with SUSY, then it is fair to say, based on the available literature, 
that much has been done in the second respect. Various methods to extract the presence
of new resonant SUSY particles have in fact been devised over the years and applied to 
several possible model realisations of a SUSY theory, particularly for the case of the 
simplest SUSY scenario, the MSSM. In constrast, much less has been done so far in the first 
respect. Primarily because the LHC is perceived not to be a precision machine, 
due to both the undefined partonic energy and the intrinsically large hadronic background, 
so that virtual effects (of order $\alpha_s$ and $\alpha$ or indeed smaller) are 
expected to be not easily discernible in the data. However, there are certain processes 
where the experimental precision is expected to eventually become comparable to the size 
of the virtual effects of NP. 

Among these processes, one can certainly list top quark production, both in double-
and single-top mode. In fact, with typical hadro-production cross sections at the LHC
of order 800 pb and 300 pb respectively (at 14 TeV) and collider luminosities that
can reach the 300 fb$^{-1}$ level, several hundred million top events will altogether be produced
during the lifetime of the CERN machine. Clearly, on the one
hand, this renders the statistical error applicable on the experimental side to typical 
top quark observables negligible. Furthermore, on the other hand, the main systematics
affecting the latter (both theoretical, coming from the Parton Distribution Function (PDF) 
dependence, and experimental, coming from the jet energy scale uncertainty) will
be understood at the percent level after the first hadronic data samples will have been 
collected and studied in detail. 

Therefore, both double- and single-top production processes at the LHC lend themselves 
to being precision physics laboratories, where, in particular, virtual effects of SUSY
(when $M_{\rm{NP}}\equiv M_{\rm{SUSY}} \gg m_t$, where $m_t$ is the top quark mass)
can possibly be extracted, in the ultimate attempt to understand the underlying dynamics
of SUSY breaking. We intend to illustrate here the potential afforded in this respect
by two single-top production processes, namely the t-channel:
\begin{equation}\label{tq}
bq\to tq'
\end{equation} 
and the s-channel:
\begin{equation}\label{tb}
q\bar q'\to t\bar b
\end{equation} 
(plus, of course, their charge conjugated channels), where $q^{(')}$ represents a light 
quark ($u,d,s$ or $c$)

The plan of the paper is as follows. The next section describes what is available
in current literature concerning single-top processes at the LHC in terms
of their higher order corrections within the MSSM. Sect.~\ref{Sec:Calc} illustrates
how we performed the calculation of the one-loop SUSY QCD and EW corrections to the
two single-top channels (\ref{tq})--(\ref{tb}). The following section presents our
numerical results. We conclude in Sect.~\ref{Sec:Conclusions}.


\section{Single-top Processes at the LHC}
\label{Sec:Processes}
Due to the their relevance for LHC physics, single- as well as double-top processes have extensively been studied in previous years,
with a twofold aim: on the one hand, in order to have a precise SM
prediction, at least at the complete NLO (and possibly next-to-NLO (NNLO) for the QCD
corrections); on the other hand, in order to investigate possible deviations from SM predictions due to the presence of NP. These two goals are
obviously related, the former being a precondition for the second: 
in order to extract meaningful information from NP processes entering at one-loop level top quark  hadro-production at the LHC, clearly, all similar SM effects should be well under control.

One-loop SM corrections to single- and double-top production in hadron-hadron collisions
have been known for some time. These include the SM QCD \cite{QCD-SM-tt} and SM EW 
\cite{EW-SM-tt} ones to $gg,q\bar q\to t\bar t$ (double-top production) and the 
corresponding ones \cite{QCD-SM-t,EW-SM-t} for $bq\to tq'$, 
$q\bar q'\to t\bar b$ and $bg\to tW^-$ (single-top production). 
Both SM QCD and EW corrections can be large, although in complementary 
energy regimes: $\sqrt s\approx m_t$ and $\sqrt s\gg m_t$, respectively. However,
they all are rather stable against variations of the factorisation and/or renormalisation
scales (these in turn quantifying the systematic uncertanties related to the unknown
two-loop, or even higher order, corrections), so that one can conceivably attempt
to investigate virtual effects of some NP, particularly of SUSY origin, induced
onto single- and double-top processes. To stay with the
MSSM, QCD and EW corrections to double-top production have been quantified in
\cite{QCD-MSSM-tt} and  \cite{EW-MSSM-tt}, respectively, whilst only part of these
are known to date for the case of single-top reactions~\cite{EW-MSSM-t}.  


Our paper is thus the first one where both MSSM QCD and EW corrections to single-top
processes inducing a `$t$ plus jet' signature in the final state are computed
(i.e., $bq\to tq'$ and $q\bar q'\to t\bar b$). It is in fact
from this perspective that we decided to postpone to another publication
the presentation of similar corrections
to the third single-top channel (i.e., $bg\to tW^-$), in the sense that the latter originates
an altogether different signature in the final state, `$t$ plus lepton and missing (transverse)
energy' or `$t$ plus two jets', depending on whether the $W^\pm$ boson produced in association
with the top quark decays leptonically or hadronically. In fact, both of the latter require
different triggers and selection procedures with respect to the former, so that, from a
phenomenological point of view, they deserve a separate treatment. 


\section{Calculation}
\label{Sec:Calc}



The values of the SM input parameters considered for the numerical evaluation of the one-loop
corrections can be found in Tab.~\ref{tab:SMinputs}. 

\begin{table}[htbp]
\centering
\begin{tabular}{@{\extracolsep{5pt}}|c|c|}
 \hline
  Coupling constants & $\alpha(M_Z)=1/127.93400652 \quad \alpha_s(M_Z)=0.1176$ \\
  \hline
  Gauge boson masses & $M_W = 80.424~{\rm{GeV}} \quad M_Z = 91.1876~{\rm{GeV}}$ \\
  \hline
  Heavy Quark Masses & $m_t = 170.9~{\rm{GeV}} \quad m_b^{\overline{MS}}(m_b) =
4.2~{\rm{GeV}}$ \\
  \hline
\end{tabular}
\caption{Numerical values of SM inputs.}
\label{tab:SMinputs}
\end{table}

We used  the CTEQ6 2006 PDF set~\cite{Pumplin:2002vw}, in particular the CTEQ6L fit, which is LO in
QCD, choosing renormalisation and factorisation scales both equal to the top mass, $m_t$.

As intimated in the introduction, due to the relatively large single-top production cross sections
(in particular for the t-channel
process) and the consequent large available statistics, we reckon that it will be possible at the LHC to
perform ``precision physics'' of top samples produced singly. It is therefore a worthy
task trying to understand systematically if sizable virtual effects could modify 
the SM predictions, providing an indirect but encouraging hint of NP.
As a first step we performed an adaptive scan over both the MSSM and minimal-SUper GRAvity (mSUGRA)
parameter spaces in order
to investigate whether there exist regions with large
one-loop corrections. We have done so limitedly to the inclusive production cross sections in either channel. 

While in the MSSM case we have input the relevant parameters directly
at the EW scale, 
in the mSUGRA scenario the low energy spectra have been obtained evolving 
the input parameters from the Grand Unification Theory (GUT) scale down to the EW scale 
through the code \verb#SUSPECT#~\cite{Djouadi:2002ze}.

We have developed a dedicated {\tt C++} code to compute the one-loop corrections and, as an internal check, we have tested the
cancellation of Ultra-Violet (UV) divergences appearing in the loop integrations.  To quantify the one-loop corrections, we will mainly focus on $K$-factors, defined in general as the ratio NLO/LO
of a given observable, with all relevant quantities (QCD and EW coupling constants, etc.)
evaluated at the given order consistently.
The PDFs, on the other hand, have been kept at LO since there NLO effects are of SM origin, whereas we are
looking here at purely SUSY corrections, therefore using NLO QCD PDF sets would introduce
spurious SM effects from higher orders.

Guided by the inclusive results, we have then looked at differential distributions for some `benchmark points'. In particular, alongside the typical kinematical observables (invariant masses, transverse momenta, (pseudo)rapidities), due to the possibility of measuring spin-related observables in top samples (the top (anti)quark decays before hadronising in fact, thus efficiently transmitting its polarisation state onto the ensuing decay products), we have also focused 
our attention on the following observables:

\begin{itemize}
  \item the Left-Right Asymmetry ($A_{\rm LR}$);
  \item the Forward-Backward Asymmetry ($A_{\rm FB}$), only for the s-channel.
\end{itemize}

$A_{\rm LR}$ is defined as usual by the following ratio:

\begin{equation}
  A_{\rm LR} = \frac{\sigma_{pp\to t_L + X}-\sigma_{pp\to t_R + X}}{\sigma_{pp\to t_L + X}+\sigma_{pp\to
t_R + X}}.
\end{equation}

$A_{\rm FB}$  deserves further details, since the definition of ``Forward'' and
``Backward'' regions
in a $pp$ collider is not straightforward. In the s-channel process the initial state can
essentially be
$u\bar{d}$ or $c\bar{s}$.
In the first case $u$ is a valence quark, and thus the momentum fraction it carries is much bigger
than that of
the $\bar{d}$ sea-quark.
It is then possible to reconstruct here the direction of the incoming valence quark just by identifying
the direction
of the boosted top in the final state.
This is a statistical process, since s-channel single-top production can be initiated by two
sea-quarks too,
such
as in the $c\bar{s}$ case, and the weight of the various events is given by the relevant PDF (hereafter denoted as $P$).
The definition of $A_{\rm FB}$ is then given by the following
relation~\cite{Barger:1986hd}:

\begin{eqnarray}
  A_{\rm FB}&=&\frac{1}{\int dx_1 dx_2 \sum_{q=u,d,c,s} (P_q(x_1,\mu)P_{\bar
      q}(x_2,\mu) + P_{\bar q}(x_1,\mu)P_q(x_2,\mu)} \times \\ 
&& \int \left[ dx_1 dx_2 {\hat\sigma}(pp\overset{u \bar d}{\rightarrow}t
      \bar b +X) \right.  \\ 
&& \left. (P_u(x_1,\mu)P_{\bar d}(x_2,\mu) - P_{\bar d}(x_1,\mu)P_u(x_2,\mu)) {\rm
  sign}(x_1-x_2) \right]
\end{eqnarray}
where $\mu$ is the factorisation/renormalisation scale.

We also have computed differential distributions of these quantities over the invariant mass of the
final state, the
transverse momentum of the top as well as  the rapidity of both the top and the light quark in the final state.

Coming back to the scans, in the general MSSM scenario,
we have explored over all soft parameters involved in the generation of the spectra,
while in the mSUGRA scenario we have scanned over the four standard mSUGRA input parameters.
The scans have been performed interfacing the tool \verb#adScan#\cite{Brein:2004kh} to our code.

\begin{figure}[t]
\centering
\epsfig{file=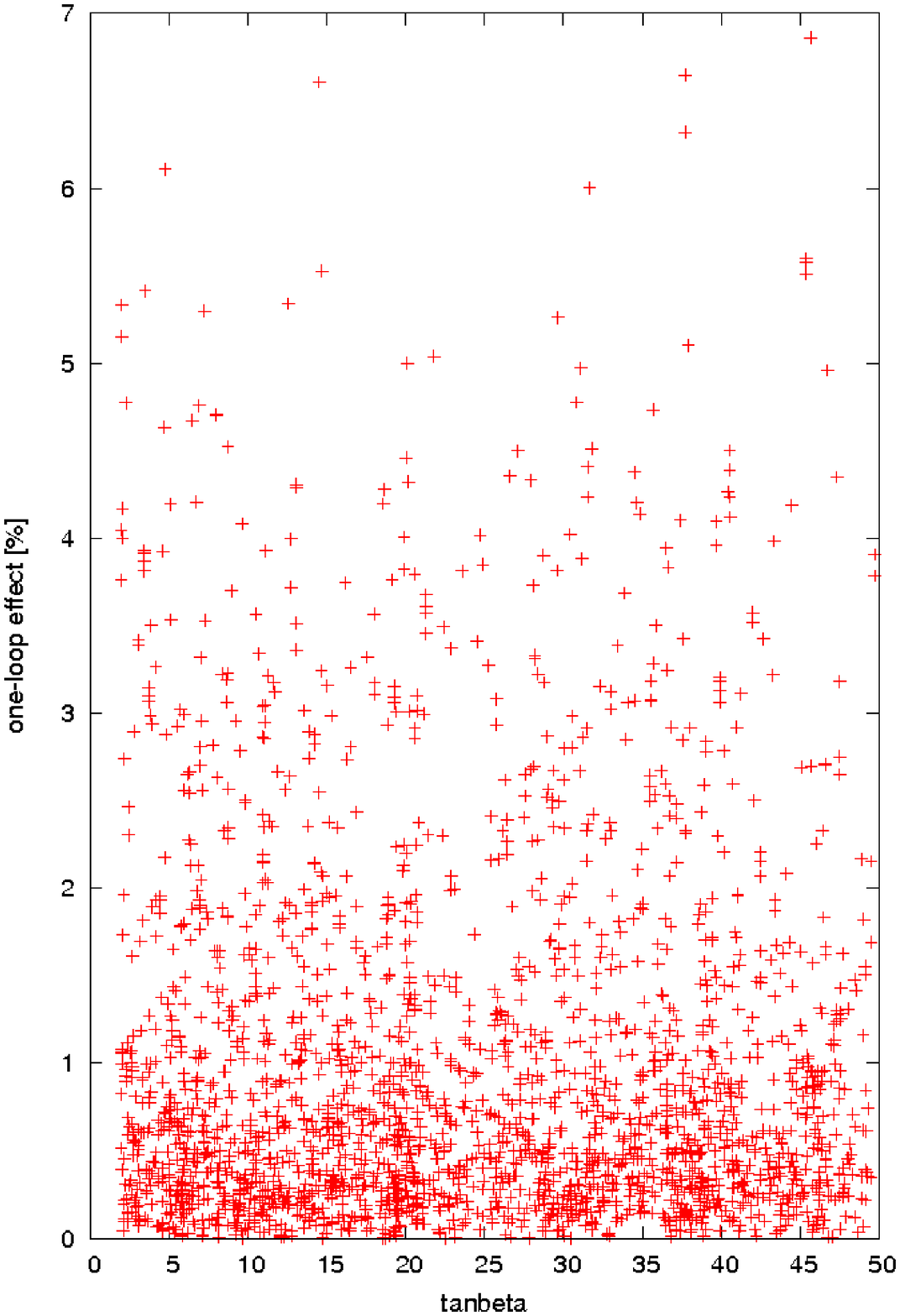,
width=0.3\textwidth, angle=0}\hfill
\epsfig{file=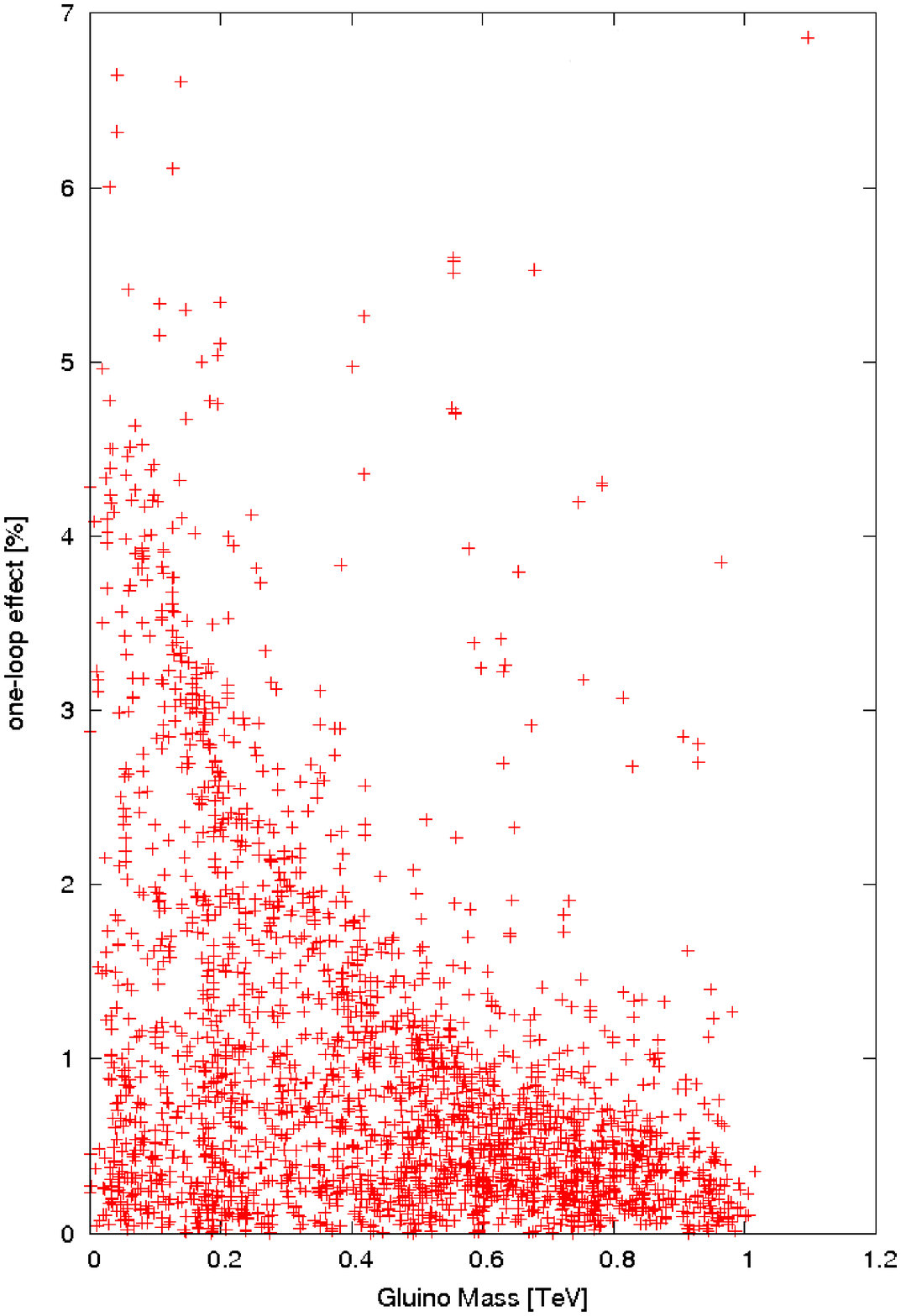,
width=0.3\textwidth, angle=0}\hfill
\epsfig{file=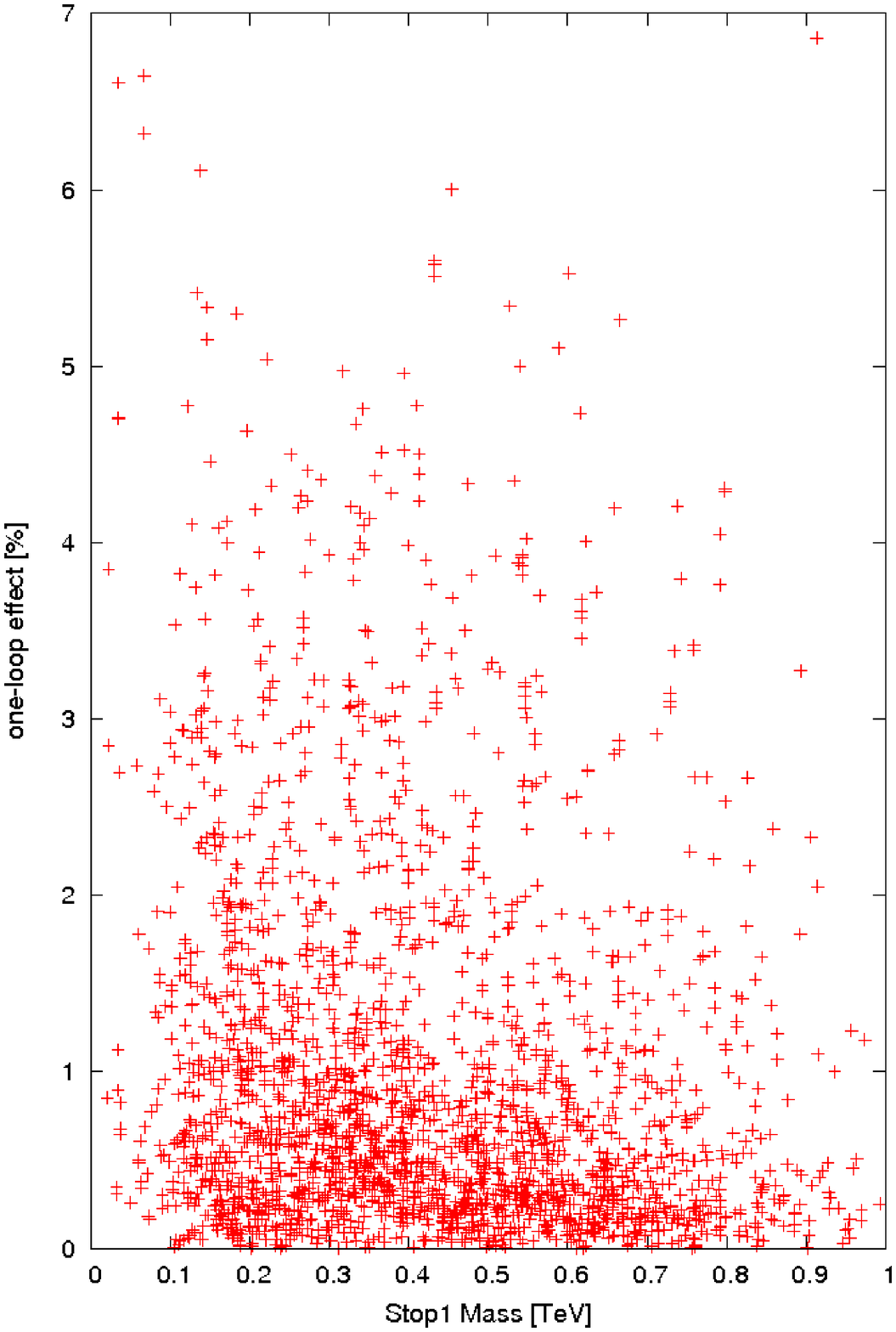,
width=0.3\textwidth, angle=0}\\[35pt]
\epsfig{file=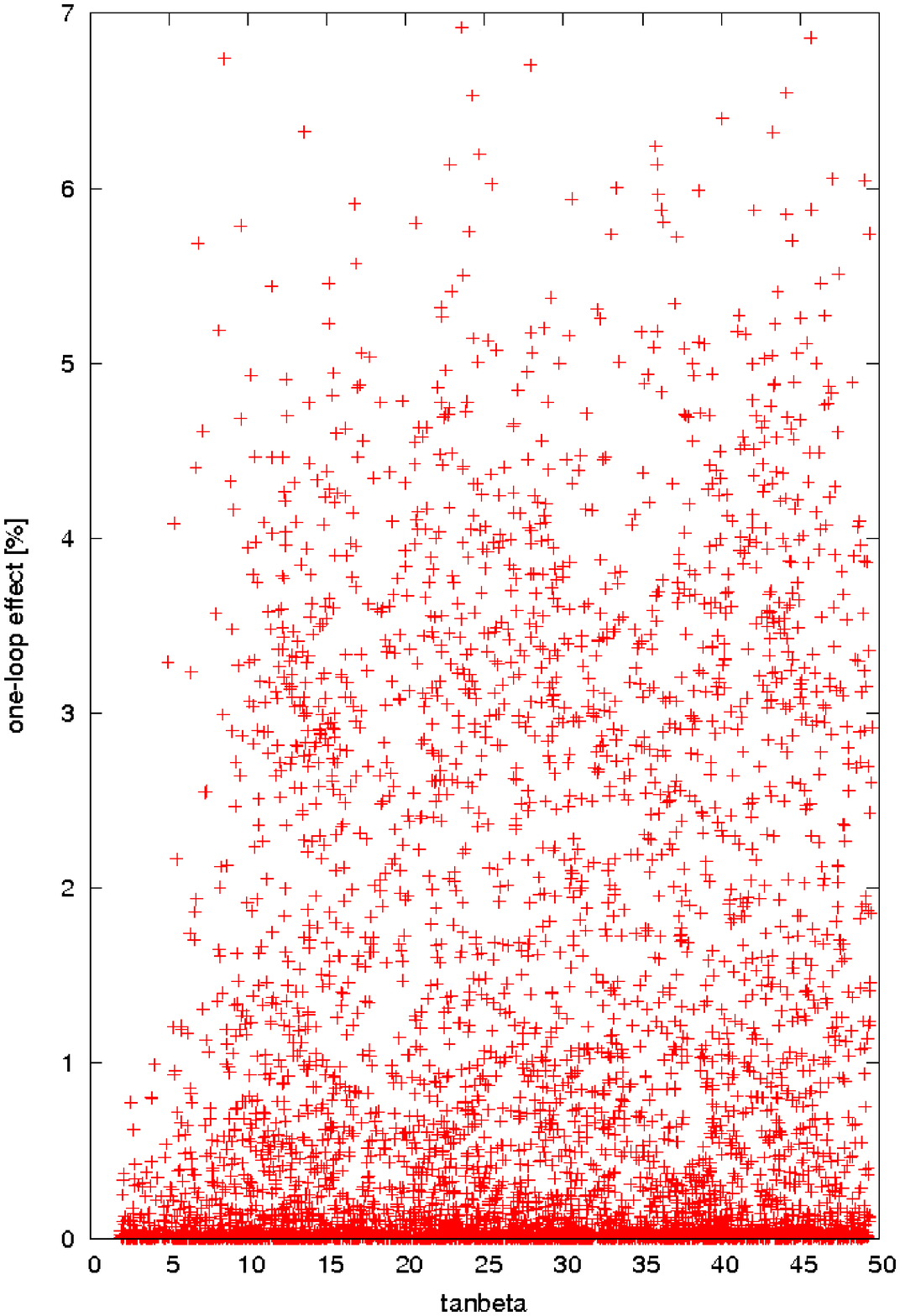,
width=0.3\textwidth, angle=0}\hfill
\epsfig{file=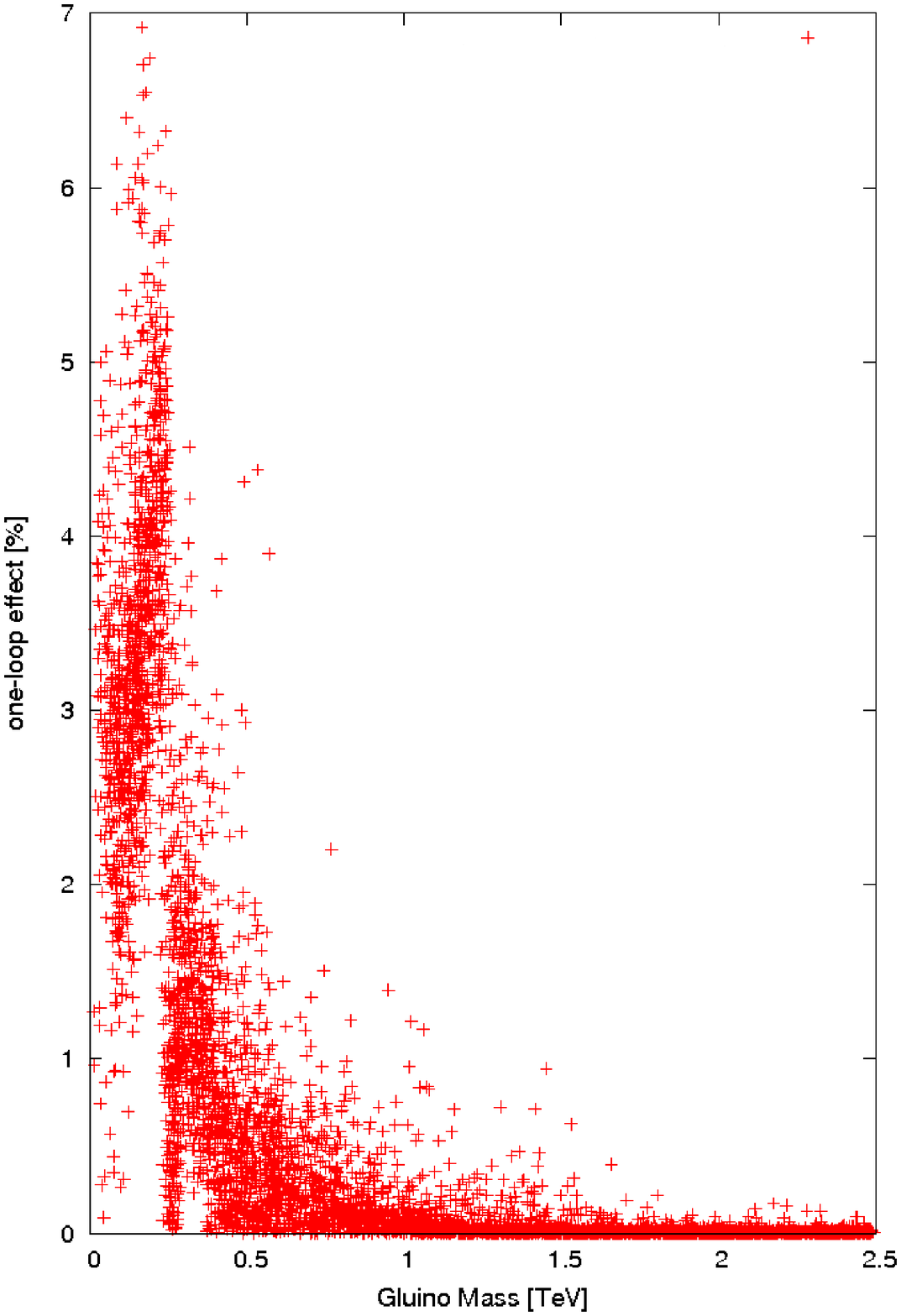,
width=0.3\textwidth, angle=0}\hfill
\epsfig{file=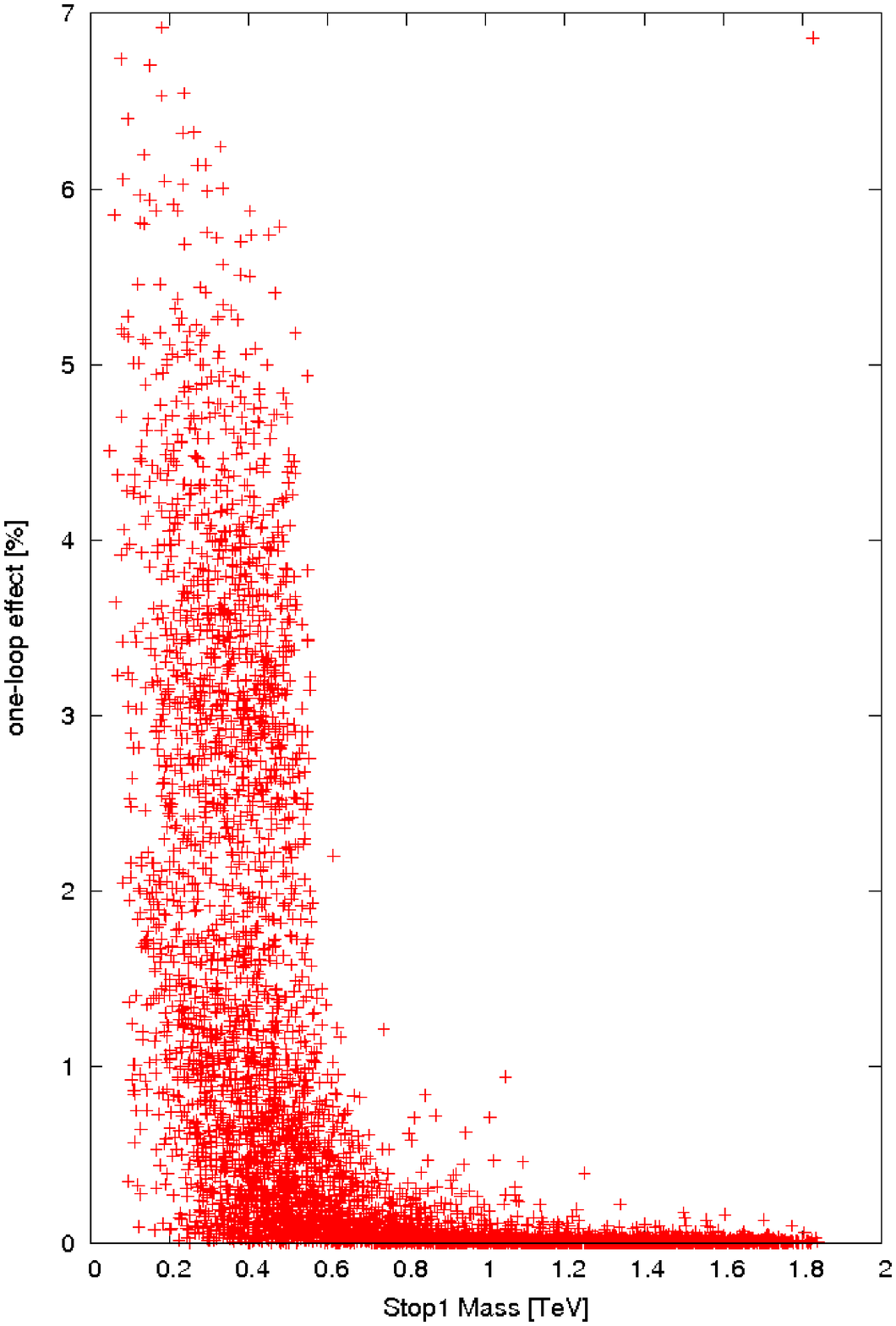,
width=0.3\textwidth, angle=0}\\[35pt]
\caption{Scan over three relevant SUSY parameters ($\tan\beta$, $m_{\tilde g}$ and
$m_{\tilde t}$) in the context of the MSSM (upper panels) and mSUGRA (lower panels) to search
for maxima of one-loop effects.}
\label{fig:AdScan}
\end{figure}

The main result of this preliminary analysis, both for MSSM and mSUGRA, is that the relative effect
of the
SUSY one-loop corrections seems to be rather small over the whole range of
the independent parameters. Moreover, the effects show a very smooth
and mild dependence on the parameters.
A meaningful sample of our results is shown in Fig.~\ref{fig:AdScan}. Each point in the diagrams
corresponds to a full computation of NLO effects for a point in the parameter space, plotted
versus a low energy parameter, such as a mass or mixing angle evaluated at the EW scale. We
have chosen to show just three of them, corresponding to the parameters which are expected to affect
most heavily the NLO corrections: $m_{\tilde g}$ (the gluino mass), $\tan\beta$
(the ratio of the two doublet Higgs vacuum expectation values) and $m_{\tilde t_1}$ (the lightest
stop mass).

From the $\tan\beta$ plot it is easy to see that the density of the points is larger for very
small values of the corrections and rapidly decreases for larger values of the effect. Furthermore,
and  possibly contrary to naive expectations, the density of the points is almost $\tan\beta$
independent. In essence, for the largest part of the parameters space, the corrections are very
small, below the 2\% level, while it is unlikely to find parameters configurations leading
to corrections as high as 5\%, and this result holds independently of $\tan
\beta$ (i.e., there are no visible regions in the $\tan\beta$ range where one-loop
effects have a maximum).  Trivially, the plots of the corrections as
a function of the gluino or stop masses simply state that it is more
likely to find ``larger'' ($6\%$ or so at the most) corrections in parameters regions where these
masses are small.

Despite the results obtained from the scan are never very large, it should be noted that these refer
to inclusive corrections only.
However, while scanning over the MSSM and mSUGRA parameter spaces,
we have also tested for the size of SUSY QCD and EW corrections to
differential distributions (including asymmetries) and found them to
be rather large, albeit in limited (and at times disfavoured) regions of phase space,
thereby explaining their smallness in the inclusive results. In order to illustrate the typical
pattern emerging at differential level, we have defined  a couple of  representative benchmark
mSUGRA points, whose parameters are given in Tab.~\ref{tab:bench}.

\begin{table}[htbp]
\centering
 \begin{tabular}{|c|ccccc|}
 \hline
 ~mSUGRA scenario~ & $\quad m_0 \quad$ & $\quad m_{1/2} \quad$ & $\quad A_0 \quad$ & $\quad \tan\beta \quad$ & $\quad \textrm{sign } \mu \quad$  \\
 \hline
 LS2         & 300 & 150 & $-500$ & 50 & + \\
 SPS1a       & 100 & 250 & $-100$ & 10 & + \\
 \hline
 \end{tabular}
\caption{Input parameters for the mSUGRA benchmark points (all values with mass dimension are in GeV).}
\label{tab:bench}
\end{table}

\begin{figure}[ht]
\centering
\epsfig{file=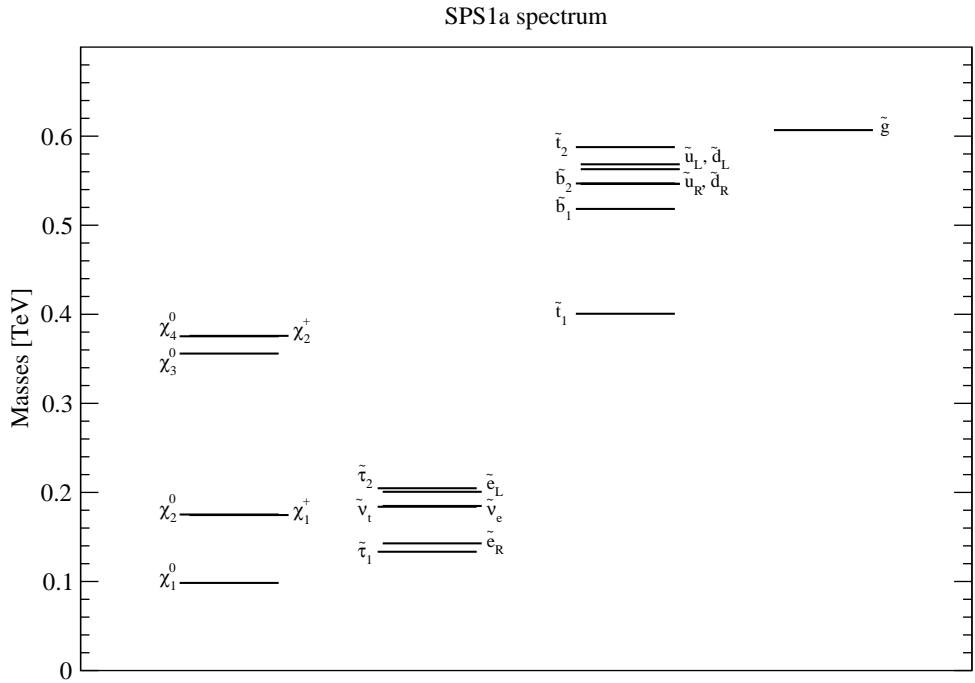, width=0.8\textwidth, angle=0}\\[50pt]
\epsfig{file=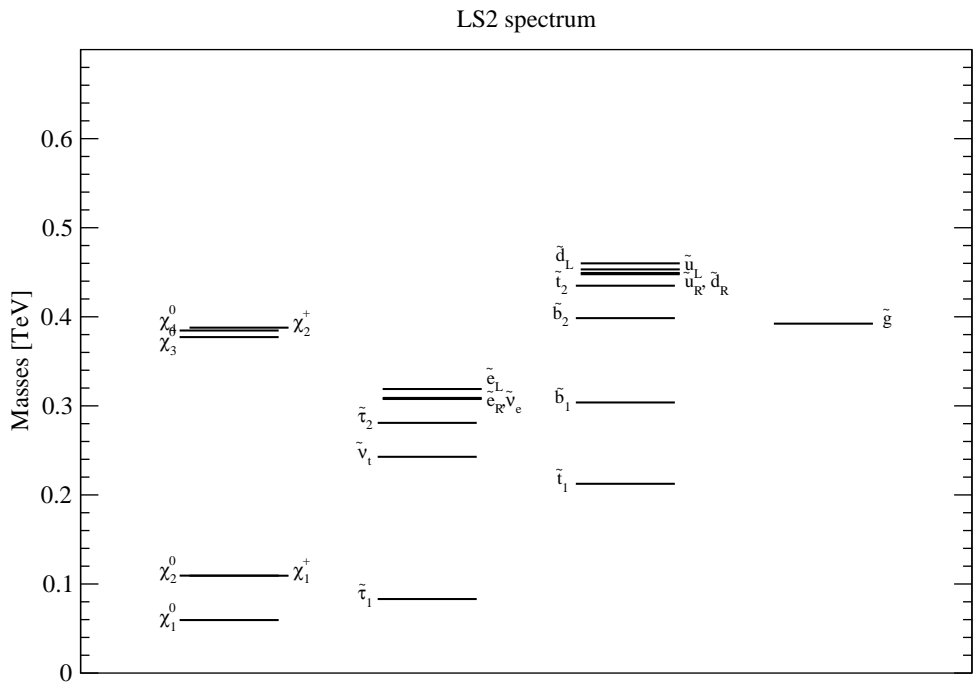, width=0.8\textwidth, angle=0}
\caption{The SPS1a and LS2 mass spectra.}
\label{fig:spectra}
\end{figure}

The SPS1a point~\cite{Allanach:2002nj} is the typical mSUGRA
``standard candle'' while the LS2 point has been introduced in~\cite{Beccaria:2006ir} and it is
characterised by an interestingly light spectrum (yet compliant with current experimental and theoretical constraints), which -- if realised in Nature -- could prompt for easily detectable signals at the LHC. More details about the two spectra can be found in
Fig.~\ref{fig:spectra}.





\vskip 1cm


\section{Results}
\label{Sec:Results}


It has recently been scheduled a long LHC run at 7 TeV, which will probably last two years and will
hopefully lead directly to the final expected energy of 14 TeV.
In view of this recent development, we present our results for both such planned energies. 
%
Our numerical results will be illustrated in turn in the following two sub-sections. Notice that we will not show all the differential distributions analysed for the two benchmark points, rather we will illustrate a meaningful sample of these (and only in the 14 TeV case) and focus on the one-loop corrections they receive.
We will also show the contribution of the SUSY EW and QCD parts separately, in some instances at least,
to better understand the features of the underlying dynamics: like, e.g., dominant contributions, asymptotic behaviours and threshold effects.

\subsection{LHC at 7 TeV}

The total cross section and asymmetries for t- and s-channel are given in Tabs.~\ref{tab:7TeV-t}
and~\ref{tab:7TeV-s} respectively.

\begin{table}[htbp]
\centering
 \begin{tabular}{|c|c|c|}
 \hline
        & $\sigma$(pb)    & L-R Asymmetry (1-$\delta$)            \\
 \hline
 Born   & 35.44           & $\delta$=4.7327$\times$10$^{-7}$  \\
 \hline
 LS2    & 35.29 ($K$=0.996) & $\delta$=4.7515$\times$10$^{-7}$ (1-$K$=1.87$\times$10$^{-9}$) \\
 SPS1   & 35.33 ($K$=0.997) & $\delta$=4.7353$\times$10$^{-7}$ (1-$K$=2.6$\times$10$^{-10}$) \\
 \hline
 \end{tabular}
\caption{Numerical results for t-channel production at 7 TeV.}
\label{tab:7TeV-t}
\end{table}

\begin{table}[htbp]
\centering
 \begin{tabular}{|c|c|c|c|}
 \hline
        & $\sigma$(pb)     & L-R Asymmetry     & F-B Asymmetry              \\
 \hline
 Born   & 2.061            & 0.6777            & 0.598049                   \\
 \hline
 LS2    & 2.086 ($K$=1.0121) & 0.6796 ($K$=1.0028) & 0.598018 ($K$=0.999948)      \\
 SPS1   & 2.062 ($K$=1.0005) & 0.6783 ($K$=1.0009) & 0.598046 ($K$=0.999995)      \\
 \hline
 \end{tabular}
\caption{Numerical results for s-channel production at 7 TeV.}
\label{tab:7TeV-s}
\end{table}

As it is possible to see, in both the two channels, the NLO corrections to the total cross sections are
opposite in sign: in the t-channel case one-loop corrections reduce the cross section, while in the
s-channel there is an enhancement, but the corrections are 1\% or less in both the benchmarks
considered (in line with the results from the mSUGRA scans).
The main difference between the two channels is of course the value of the cross section, $\sigma$. Focusing on
t-channel, which has a greater $\sigma$, we have found that, with an integrated luminosity of 1
fb$^{-1}$ the LHC will generate $\approx35000$ events per year, therefore the detection of any deviation
from the SM prediction due to the SUSY EW and QCD one-loop corrections, at least in the early stages of
the experiment, will be indeed very challenging.
Also $A_{\rm{LR}}$  shows different behaviours in the two channels but,
unfortunately, they are much greater in the s-channel, which, as we have just verified, has NLO
corrections well beyond observability at 7 TeV.
The same observability considerations can be applied to $A_{\rm{FB}}$ in the
s-channel.

%




\subsection{LHC at 14 TeV}

\subsubsection{The t-channel case}

The total cross section and asymmetries for this production mechanism are given in Tab.~\ref{tab:14TeV-t}.

\begin{table}[htbp]
\centering
 \begin{tabular}{|c|c|c|}
 \hline
        & $\sigma$(pb)    & L-R Asymmetry (1-$\delta$)            \\
 \hline
 Born   & 122.5           & $\delta$=4.3491$\times$10$^{-7}$  \\
 \hline
 LS2    & 122.0 ($K$=0.996) & $\delta$=4.3668$\times$10$^{-7}$ (1-$K$=1.77$\times$10$^{-9}$) \\
 SPS1   & 122.1 ($K$=0.997) & $\delta$=4.3516$\times$10$^{-7}$ (1-$K$=2.5$\times$10$^{-10}$) \\
 \hline
 \end{tabular}
\caption{Numerical results for t-channel production at 14 TeV.}
\label{tab:14TeV-t}
\end{table}

In this case the total cross section is significantly bigger than in the 7 TeV case, and therefore
we expect that it will be possible to extract much more information from experimental data. The
$K$-factor for the integrated cross section is admittedly small, less than 1\% for both benchmarks but,
with the expected luminosity of 10 fb$^{-1}$, the production of single-top through this t-channel
process will count $\sim$1M events and therefore it should be possible to detect a discrepancy in
the expected number of SM events that could be interpreted as a hint of NP and thus boost the search of a 
SUSY resonance. As far as the $A_{\rm LR}$ is concerned, the top quark will emerge almost completely left-polarised, yet the
detection of a one-loop induced discrepancy in the asymmetry will be again unlikely, the $K$-factor
being close to one part in a billion or less.

\begin{figure}[t]
\centering
\epsfig{file=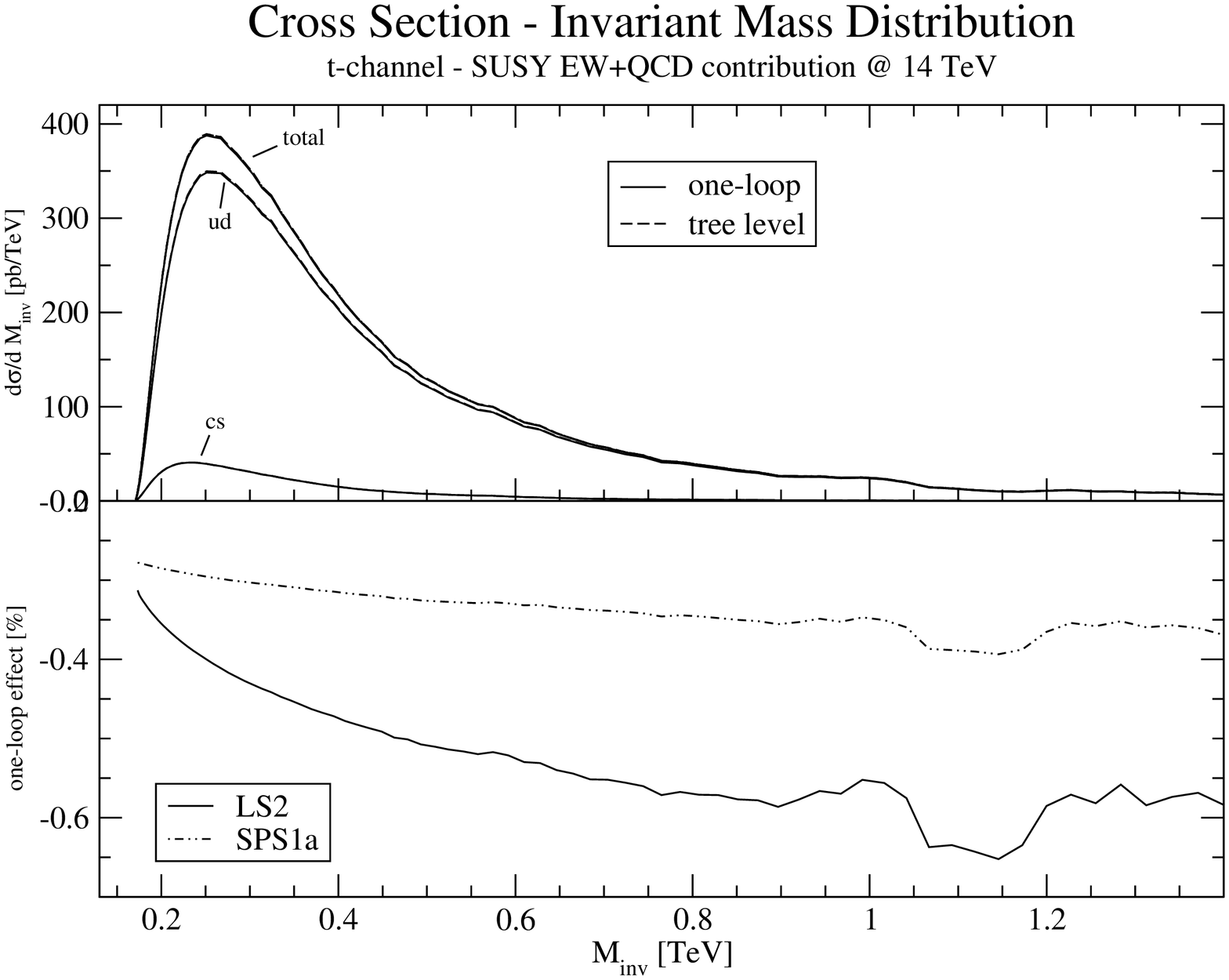,
width=0.48\textwidth, angle=0}\hfill
\epsfig{file=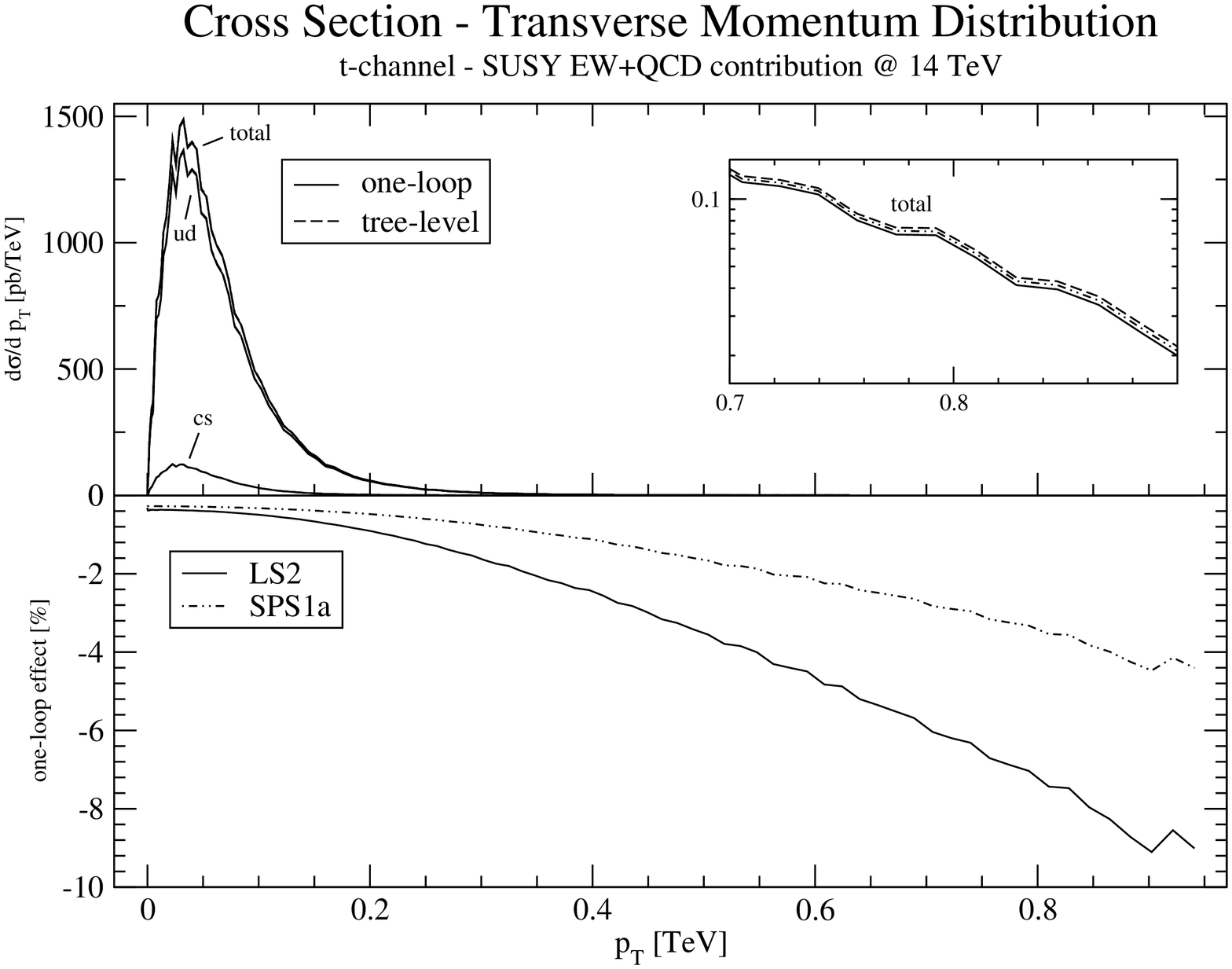,
width=0.48\textwidth, angle=0}\\[35pt]
\epsfig{file=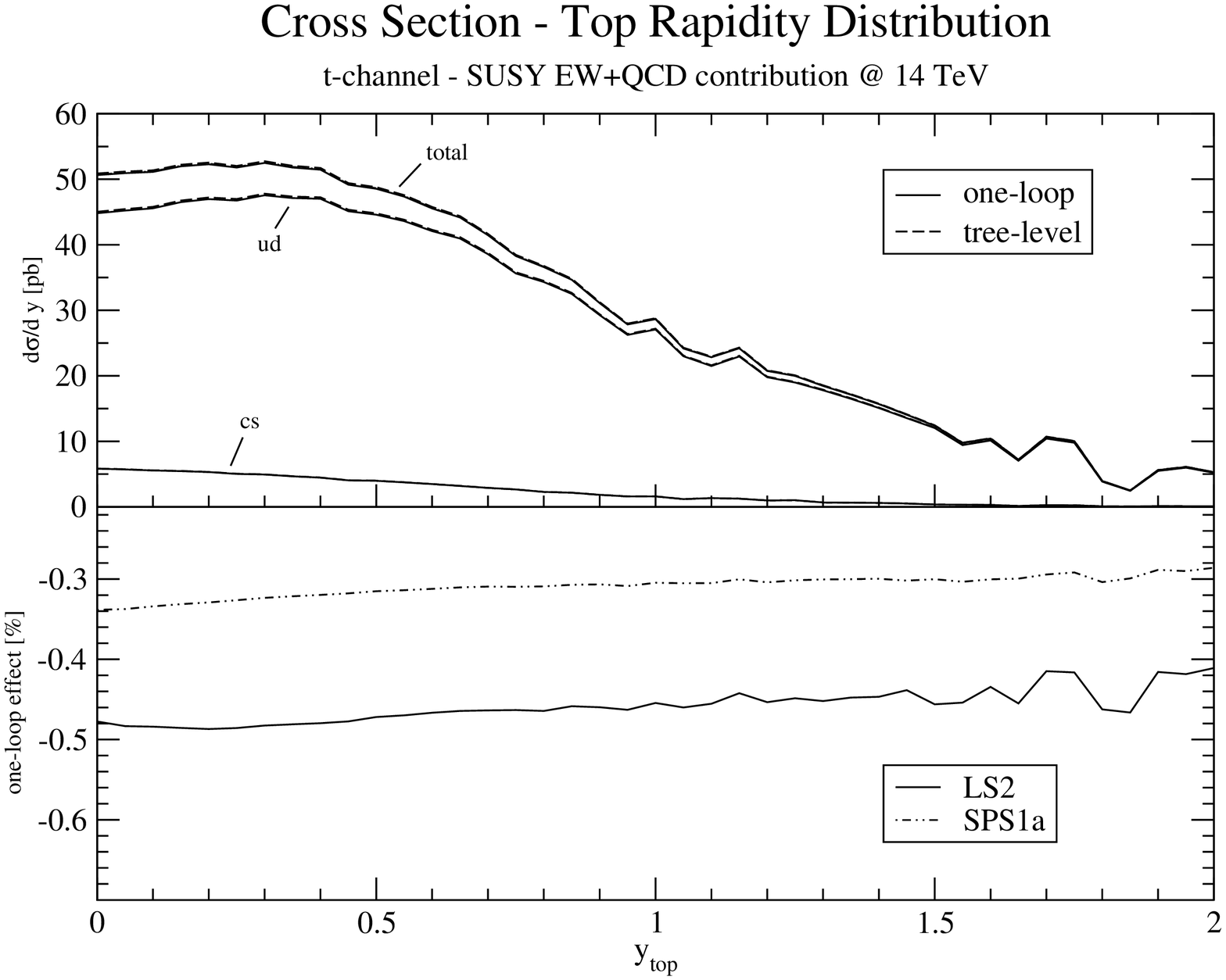,
width=0.48\textwidth, angle=0}\hfill
\epsfig{file=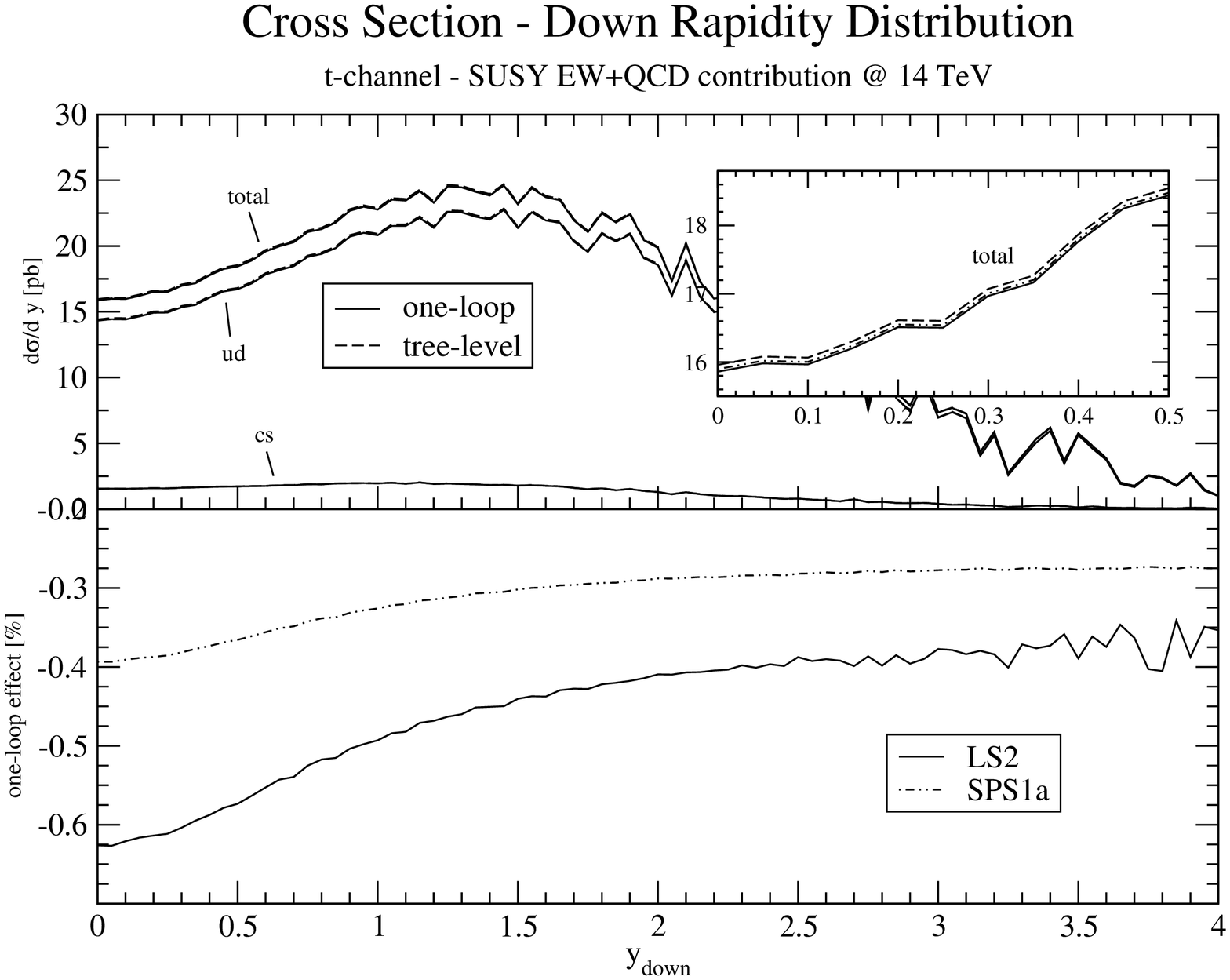,
width=0.48\textwidth, angle=0}\\[35pt]
\caption{Differential distributions of the cross section in $t$-channel at 14 TeV. In this and in
the following Figs., the small panels focus on the range where the one-loop effects are higher, and
they are shown only when the corrections from LS2 and SPS1a are distinguishable.}
\label{fig:sigma-t-channel-14TeV}
\end{figure}

The analysis of differential observables is now mandatory to better understand the origin of the
corrections to the integrated cross section and Left-Right asymmetry.
In Fig.~\ref{fig:sigma-t-channel-14TeV} we have collected our results for the differential
distributions of the cross section. It can be noticed that the corrections are always bigger in the
LS2 scenario than in the SPS1a case and that in the transverse momentum distribution they can reach the $-10\%$ limit for
high values of the top $p_t$ (transverse momentum). However, the differential cross section is quite small in this
limit, of the order of 0.01 pb/TeV, and therefore observation of large one-loop corrections for
events with high top transverse momentum is probably possible though quite difficult. It is also possible to
see that the one-loop corrections in the other differential distributions are always quite small,
below the 1\% or so limit.

\begin{figure}[t]
\centering
\epsfig{file=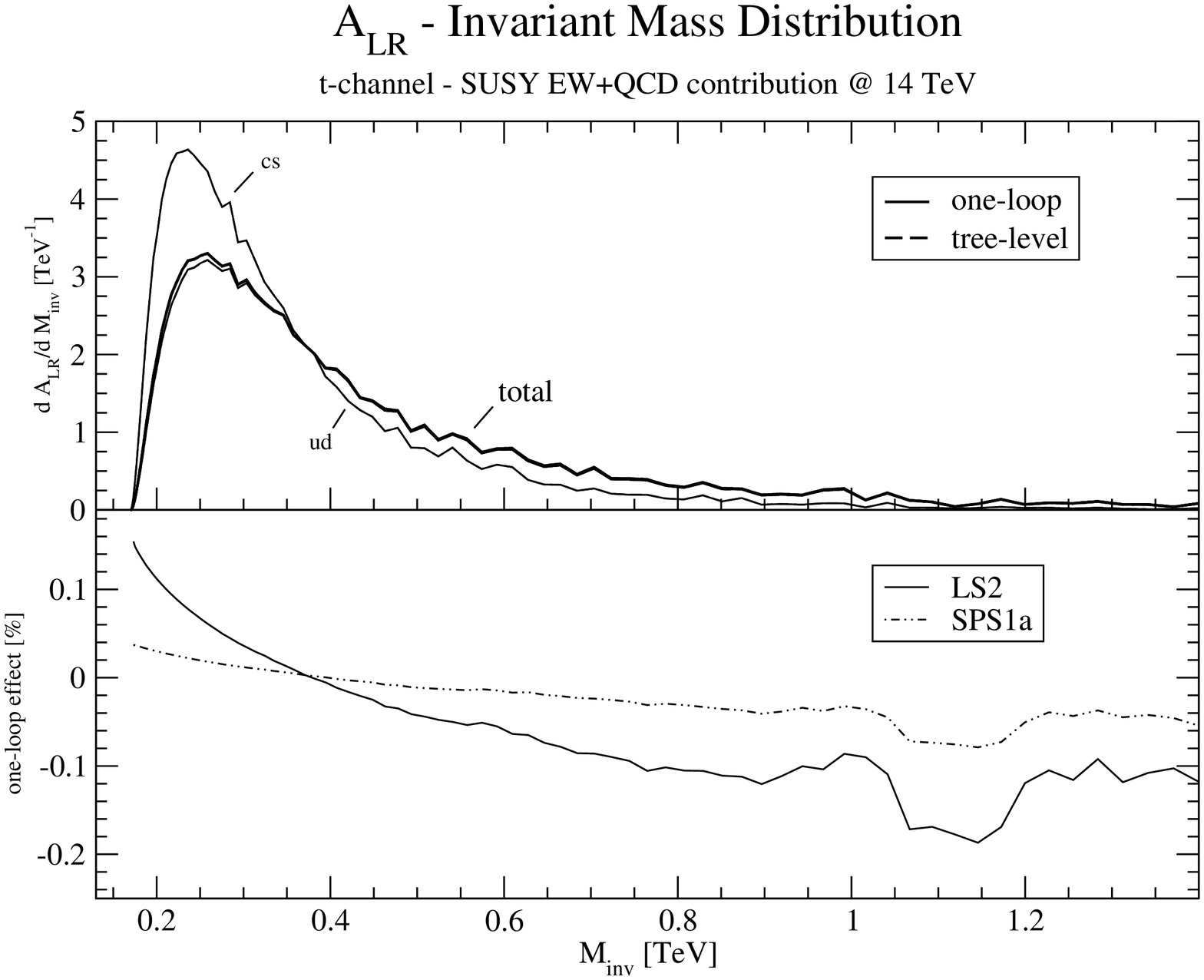,
width=0.48\textwidth, angle=0}\hfill
\epsfig{file=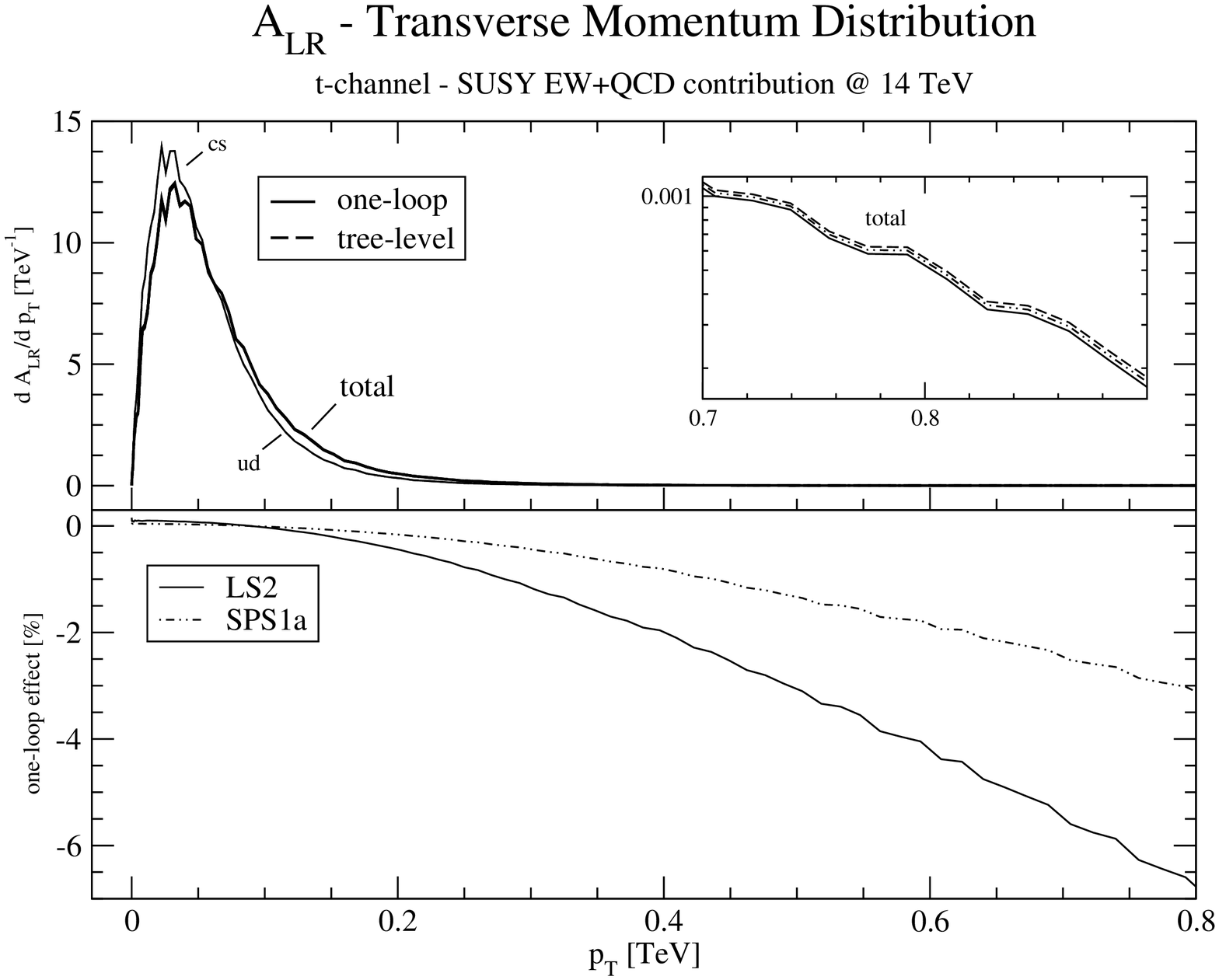,
width=0.48\textwidth, angle=0}\\[35pt]
\epsfig{file=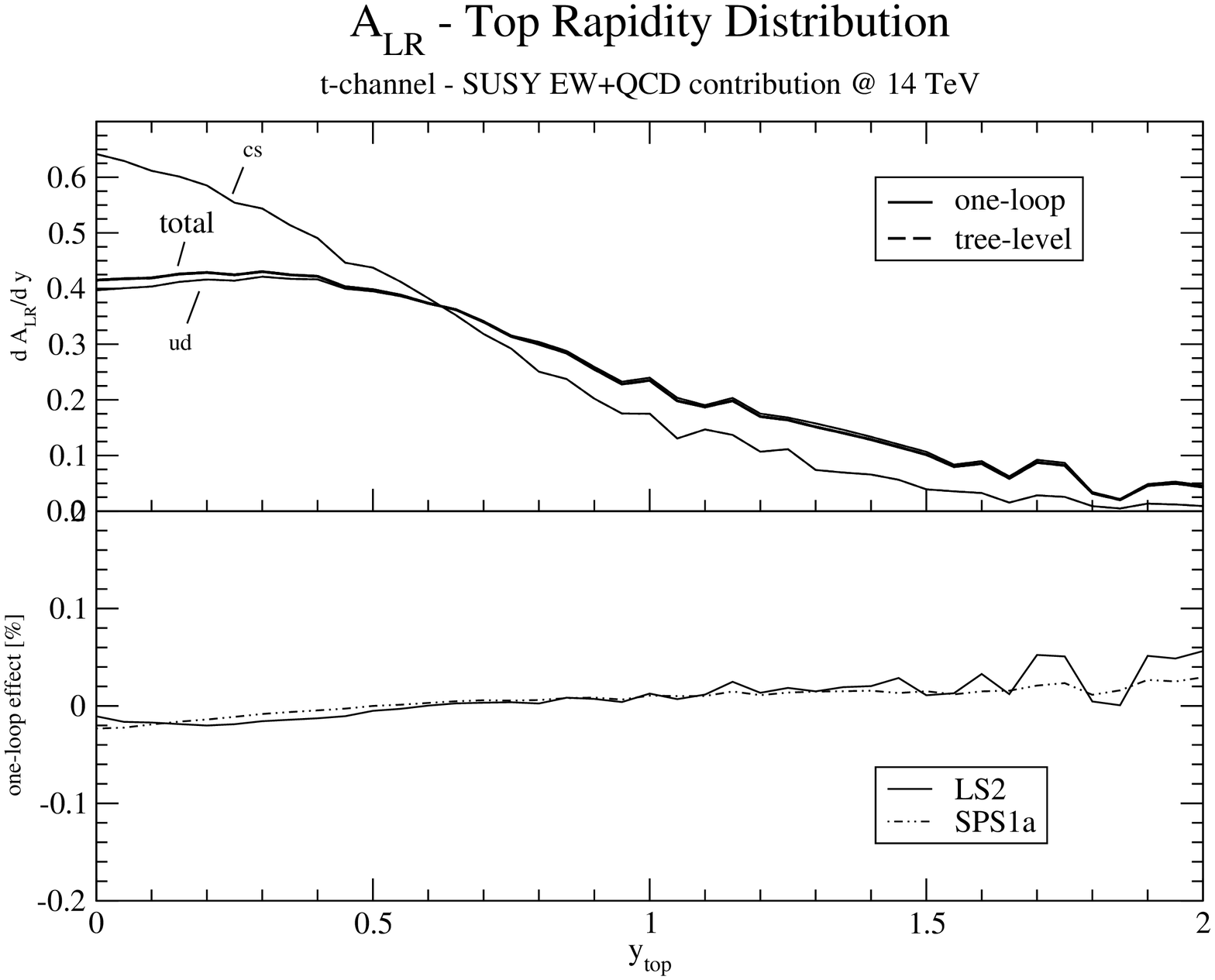,
width=0.48\textwidth, angle=0}\hfill
\epsfig{file=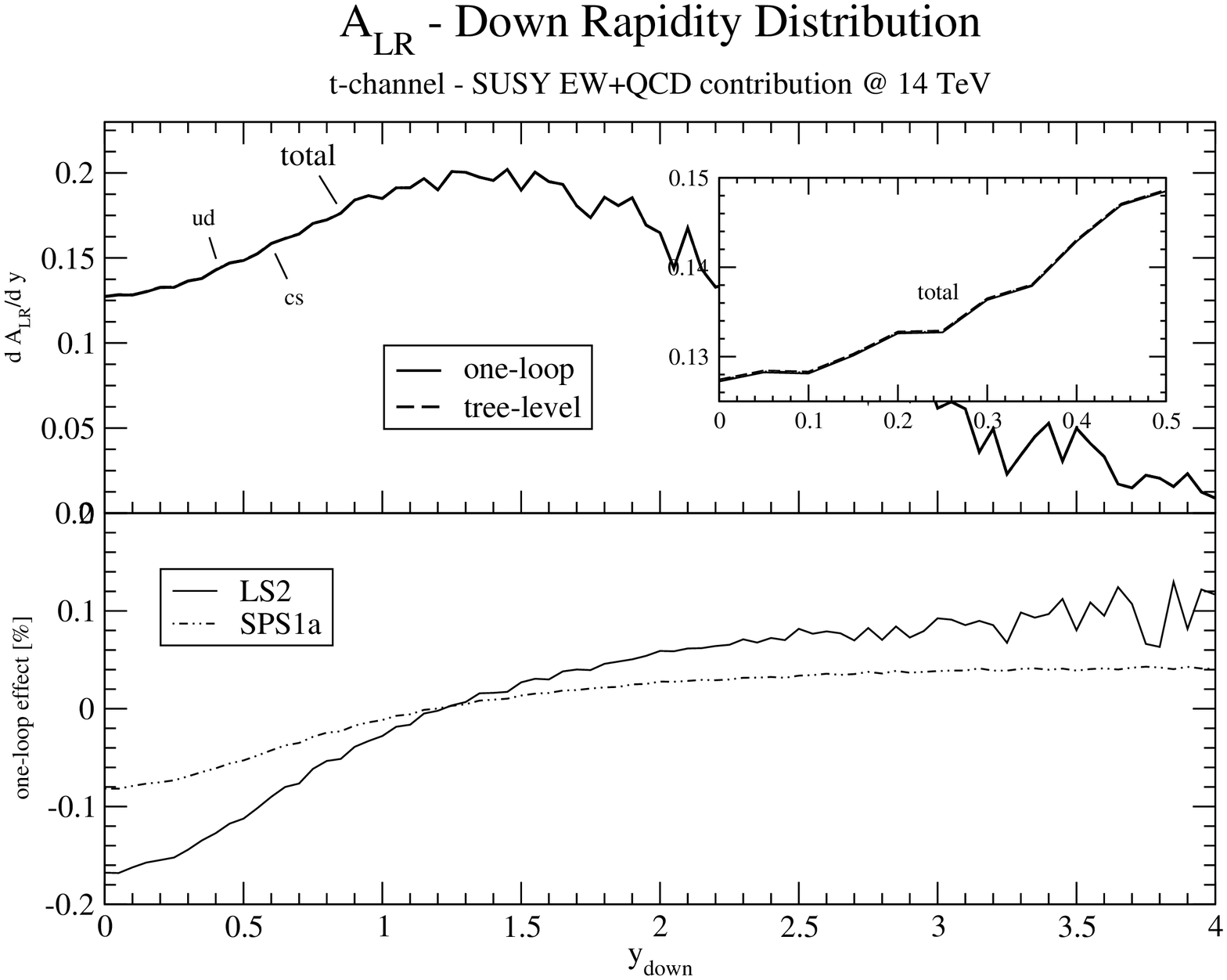,
width=0.48\textwidth, angle=0}\\[35pt]
\caption{Differential distributions of the Left-Right asymmetry in $t$-channel at 14 TeV.}
\label{fig:ALR-t-channel-14TeV}
\end{figure}

In Fig.~\ref{fig:ALR-t-channel-14TeV} the one-loop corrections to the differential distributions for
the Left-Right asymmetry are shown. Again, the corrections are generally bigger in the LS2
scenario than in the SPS1a case and can reach significant values for high top $p_t$, of $-8\%$ or so. The $p_t$ distribution has another
interesting feature: the asymmetry is bigger in the high top $p_t$ region, where, however, one-loop
corrections are much less than 1\%.

\subsubsection{The s-channel case}

The total cross section and asymmetries for this production mechanism are given in Tab.~\ref{tab:14TeV-s}.

\begin{table}[htbp]
\centering
 \begin{tabular}{|c|c|c|c|}
 \hline
        & $\sigma$(pb)     & L-R Asymmetry     & F-B Asymmetry              \\
 \hline
 Born   & 5.138            & 0.6861            & 0.53665                   \\
 \hline
 LS2    & 5.206 ($K$=1.0132) & 0.6883 ($K$=1.0032) & 0.53695 ($K$=1.00056)      \\
 SPS1   & 5.145 ($K$=1.0014) & 0.6869 ($K$=1.0012) & 0.53676 ($K$=1.00020)      \\
 \hline
 \end{tabular}
\caption{Numerical results for s-channel production at 14 TeV.}
\label{tab:14TeV-s}
\end{table}

Single-top events from s-channel at the LHC will be significantly more than at the Tevatron,
and this will allow a much more precise analysis of the process and of deviations to related
observables due to new physics.
The inclusive one-loop corrections we have obtained are at most of $\sim$1\% in the LS2 scenario,
but maybe enough to
allow a small observable difference in the expected number of events.
In constrast, the
Left-Right asymmetry receives NLO corrections of the order of 0.1\% in both the scenarios
considered (here, unlike the general case, the effect of pure SUSY corrections is almost completely negligible).
The same result holds for the one-loop corrections to the Forward-Backward asymmetry:
the $K$-factor is very small and the effect practically undetectable with the s-channel cross
section being of some pb.

\begin{figure}[t]
\centering
\epsfig{file=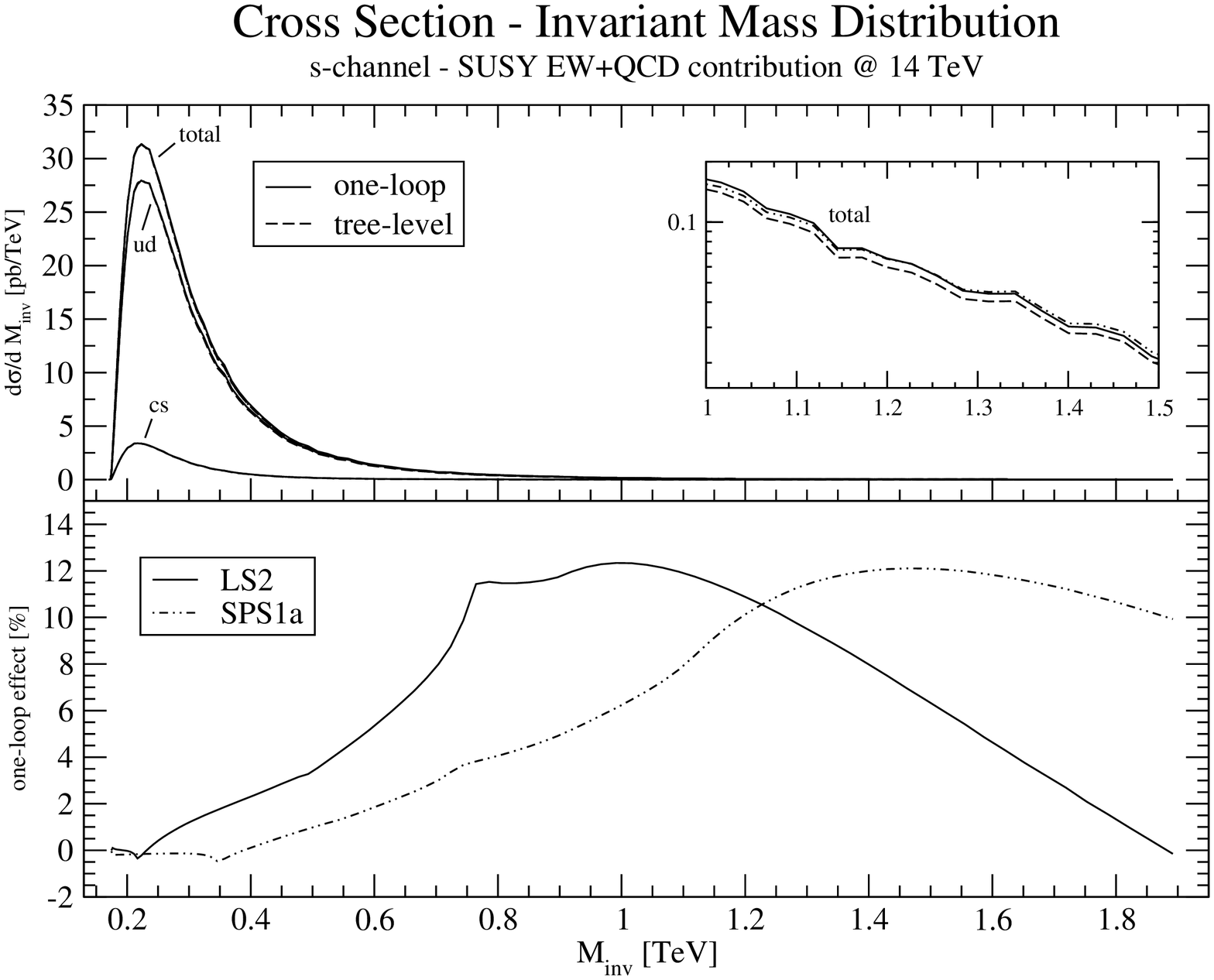,
width=0.48\textwidth, angle=0}\hfill
\epsfig{file=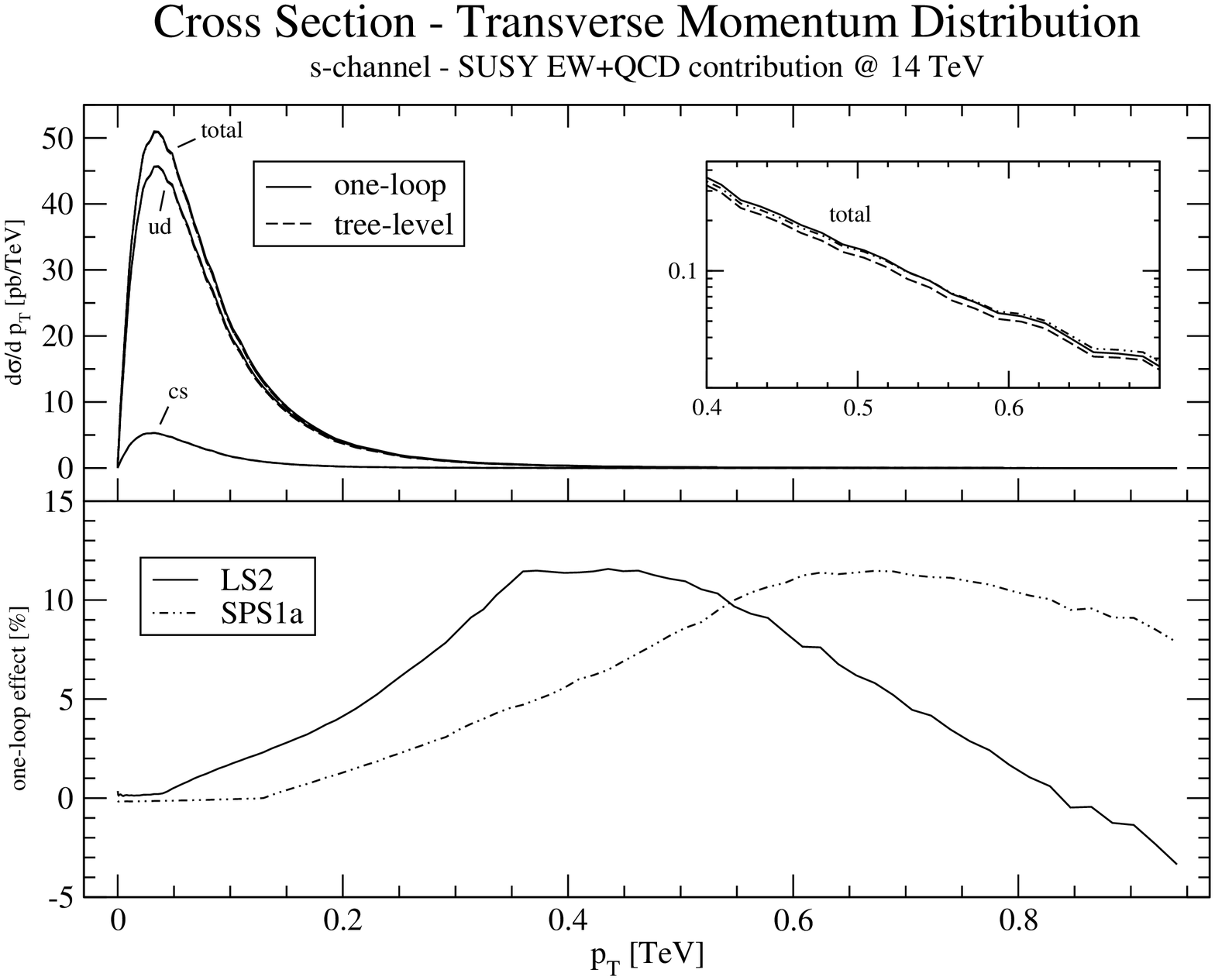,
width=0.48\textwidth, angle=0}\\[35pt]
\epsfig{file=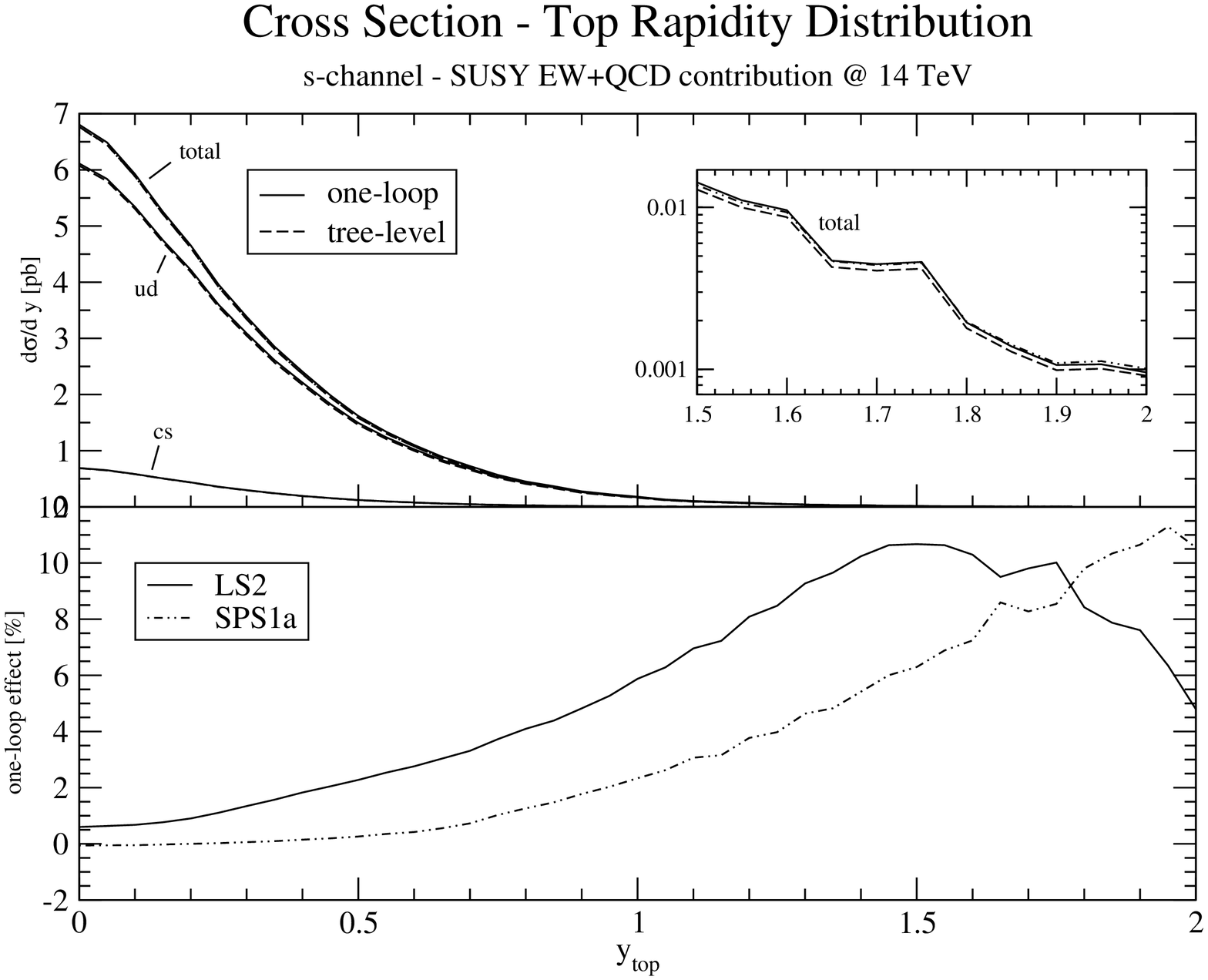,
width=0.48\textwidth, angle=0}\hfill
\epsfig{file=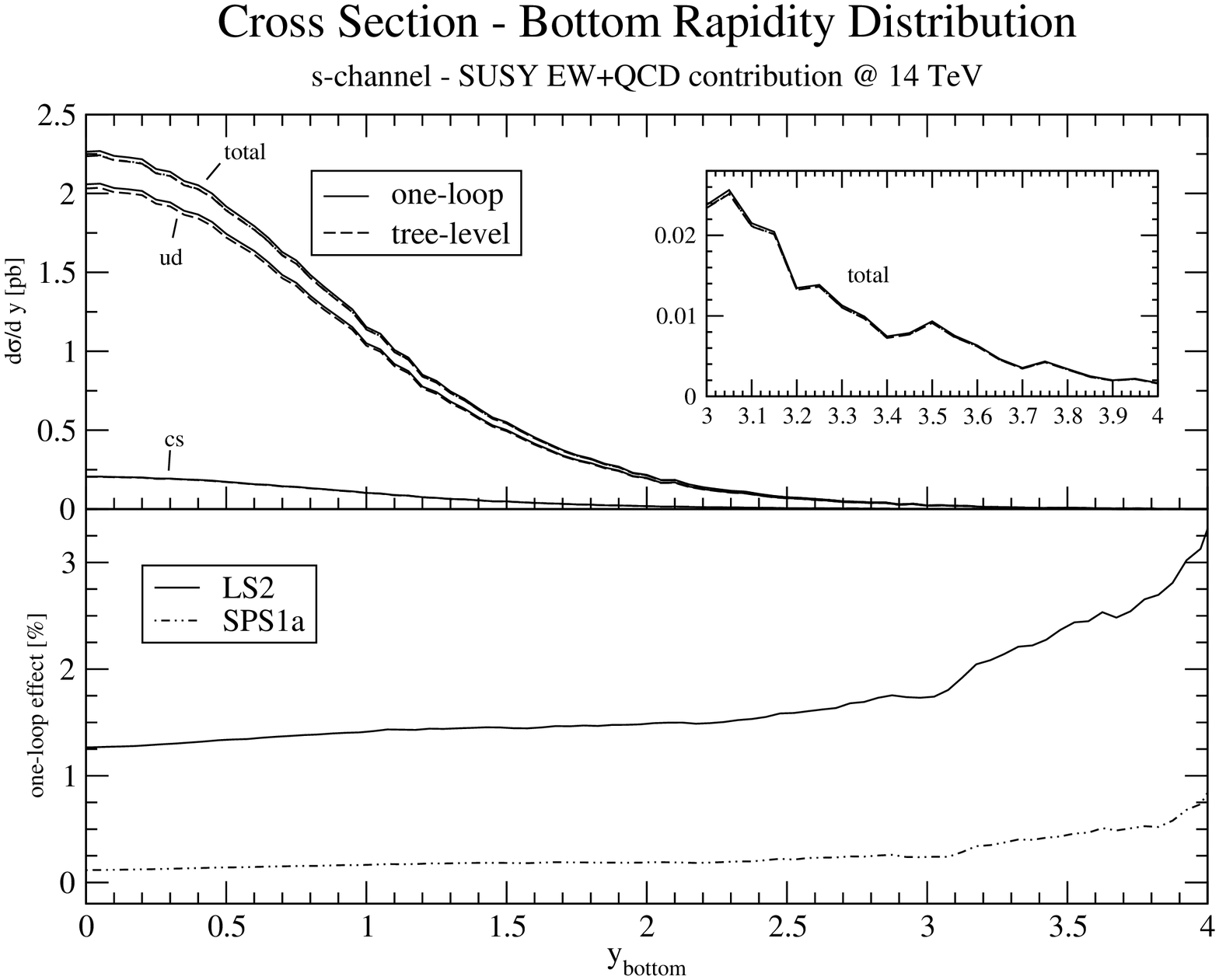,
width=0.48\textwidth, angle=0}\\[35pt]
\caption{Differential distributions of the cross section in $s$-channel at 14 TeV.}
\label{fig:sigma-s-channel-14TeV}
\end{figure}

The differential distributions for the cross section are shown in
Fig.~\ref{fig:sigma-s-channel-14TeV}. In the s-channel case there are significant differences
between the two scenarios considered: the corrections for LS2 are higher than those for SPS1a in
the regions where the contributions to the total cross section are bigger, and this is the reason
for the big difference in the $K$-factors for the integrated cross section in the two scenarios.

\begin{figure}[t]
\centering
\epsfig{file=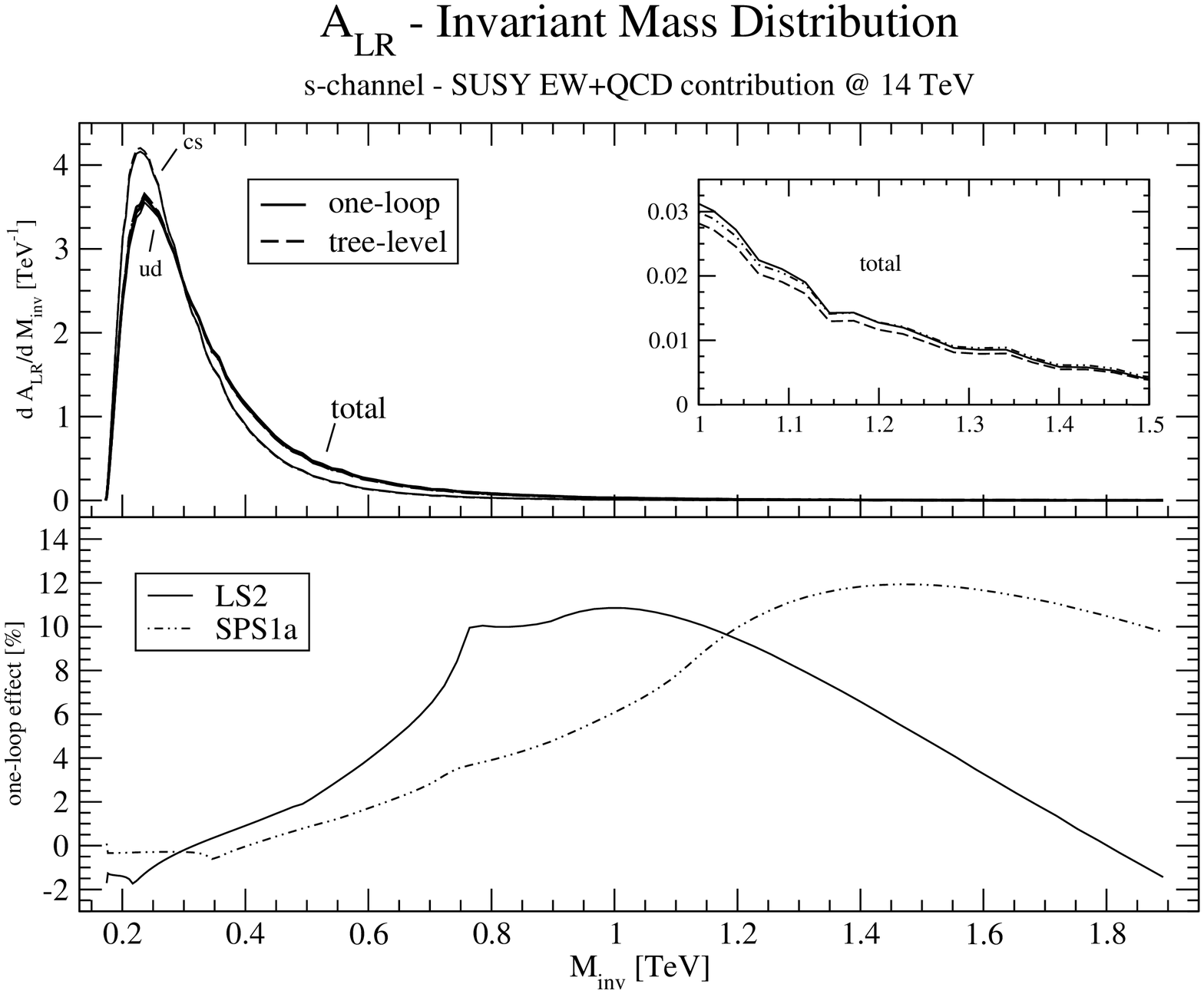,
width=0.48\textwidth, angle=0}\hfill
\epsfig{file=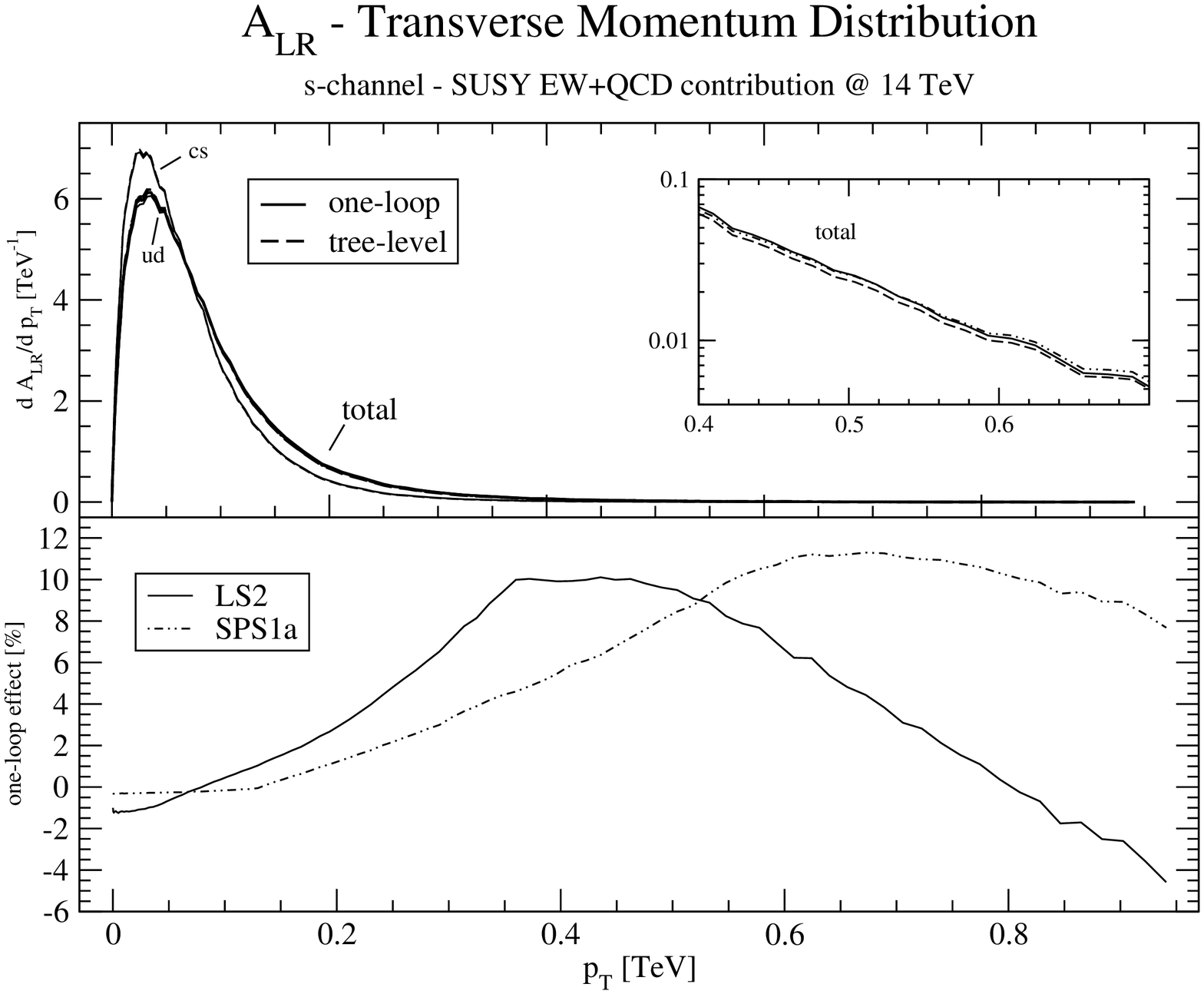,
width=0.48\textwidth, angle=0}\\[35pt]
\epsfig{file=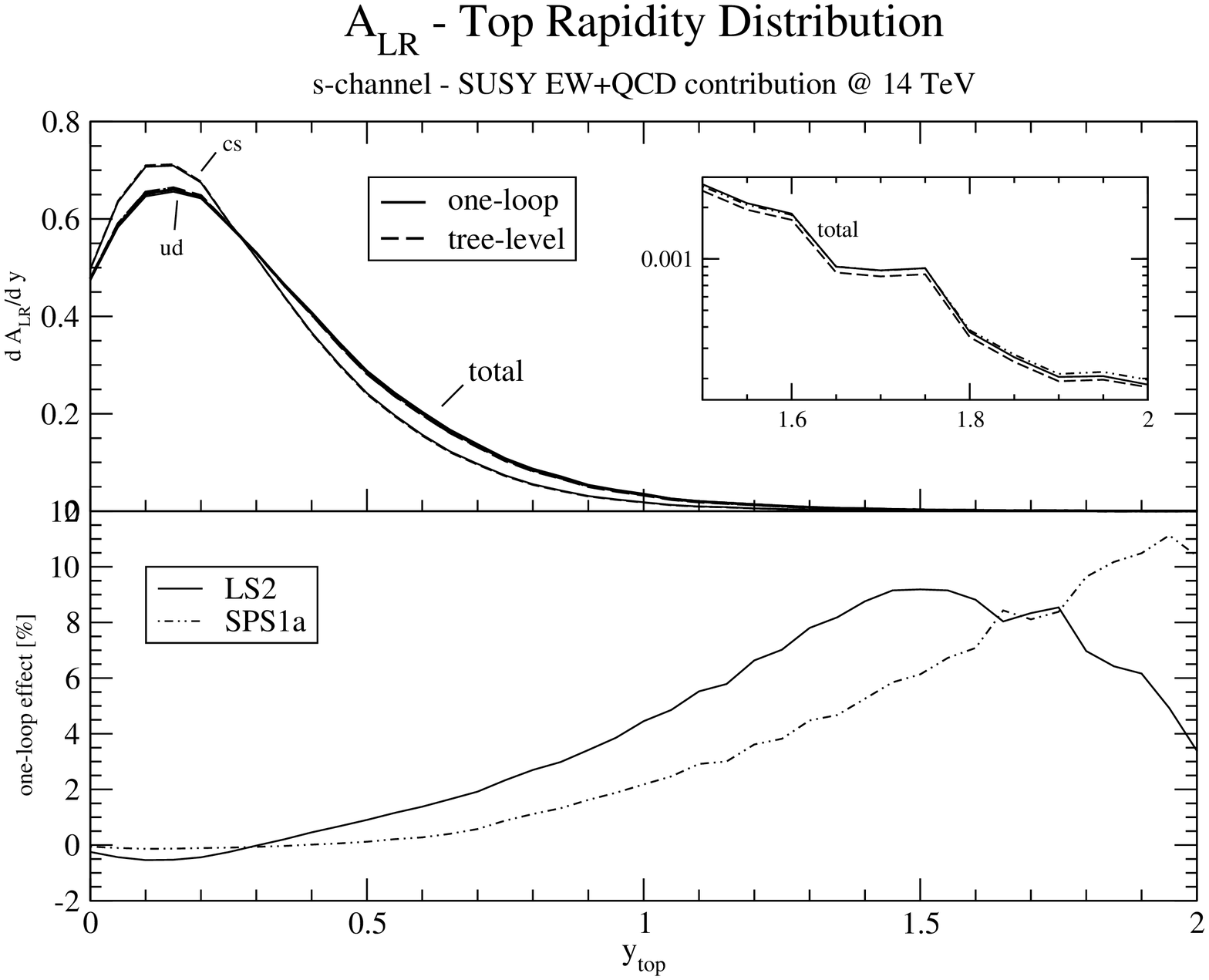,
width=0.48\textwidth, angle=0}\hfill
\epsfig{file=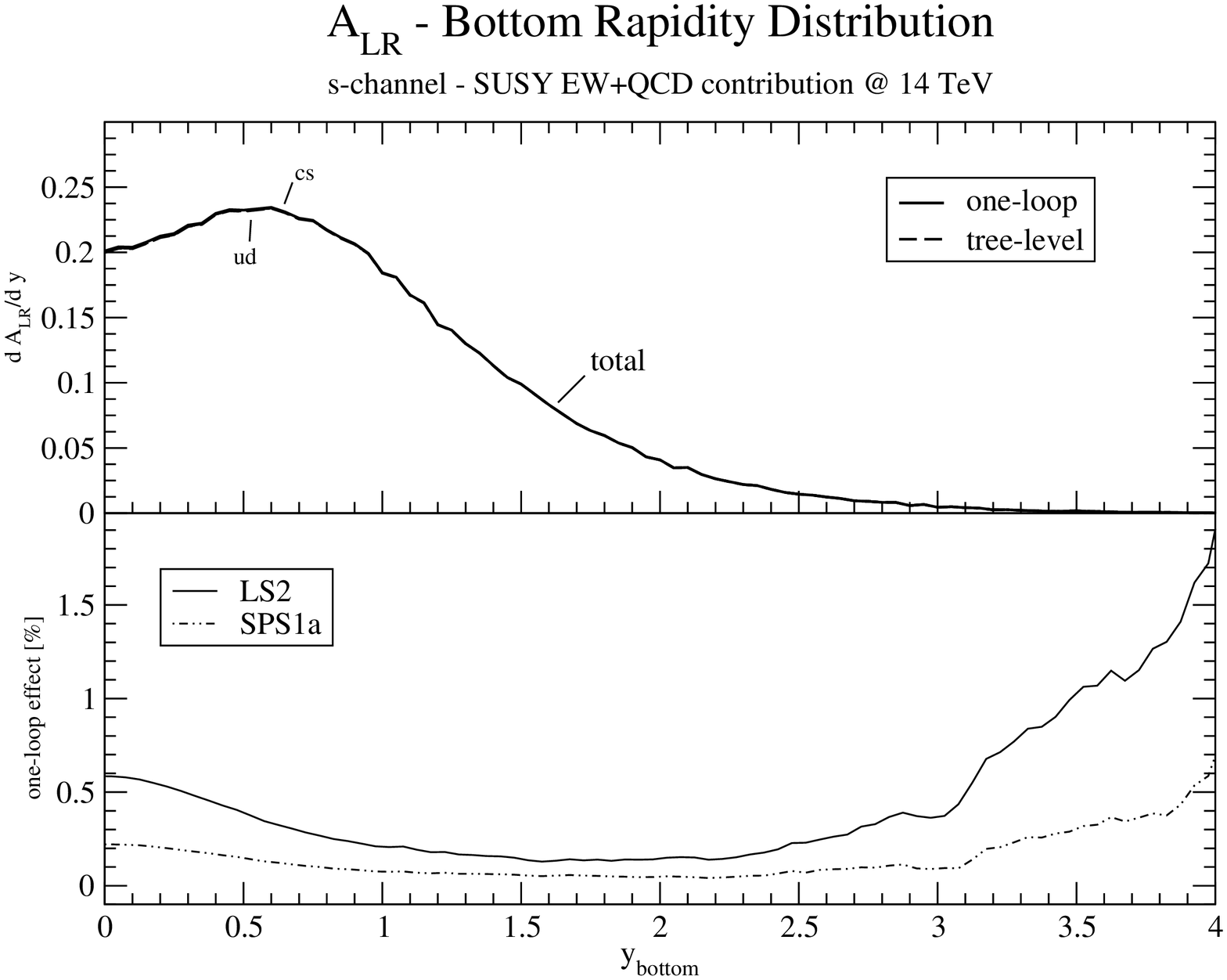,
width=0.48\textwidth, angle=0}\\[35pt]
\caption{Differential distributions of the Left-Right asymmetry  in $s$-channel at 14 TeV.}
\label{fig:ALR-s-channel-14TeV}
\end{figure}

The same feature can be observed for the Left-Right asymmetry distributions,
Fig.~\ref{fig:ALR-s-channel-14TeV}, where, however, the enhancement of the one-loop corrections from
the LS2 scenario is milder and therefore the differences in the integrated asymmetry are smaller.

\begin{figure}[t]
\centering
\epsfig{file=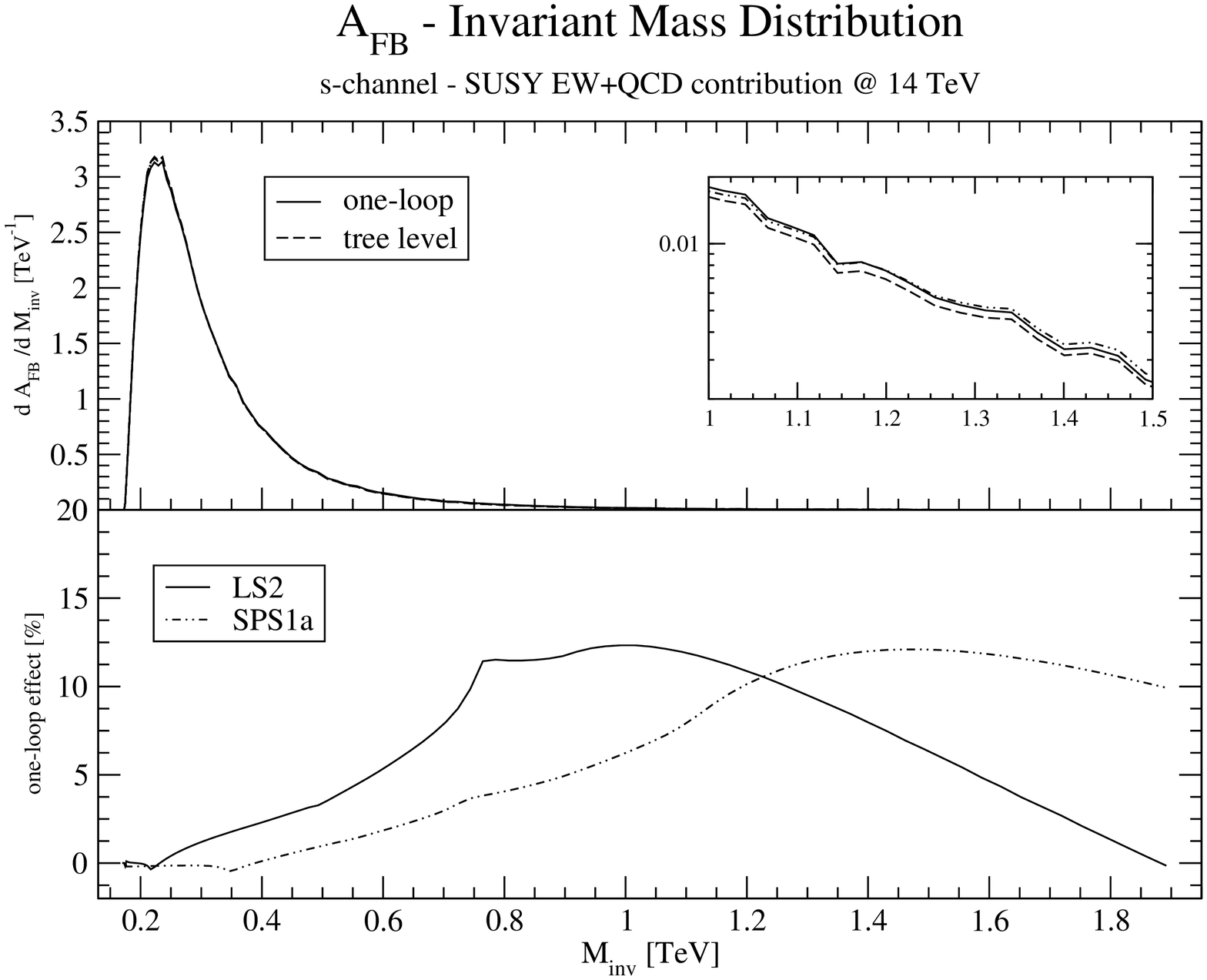,
width=0.48\textwidth, angle=0}\hfill
\epsfig{file=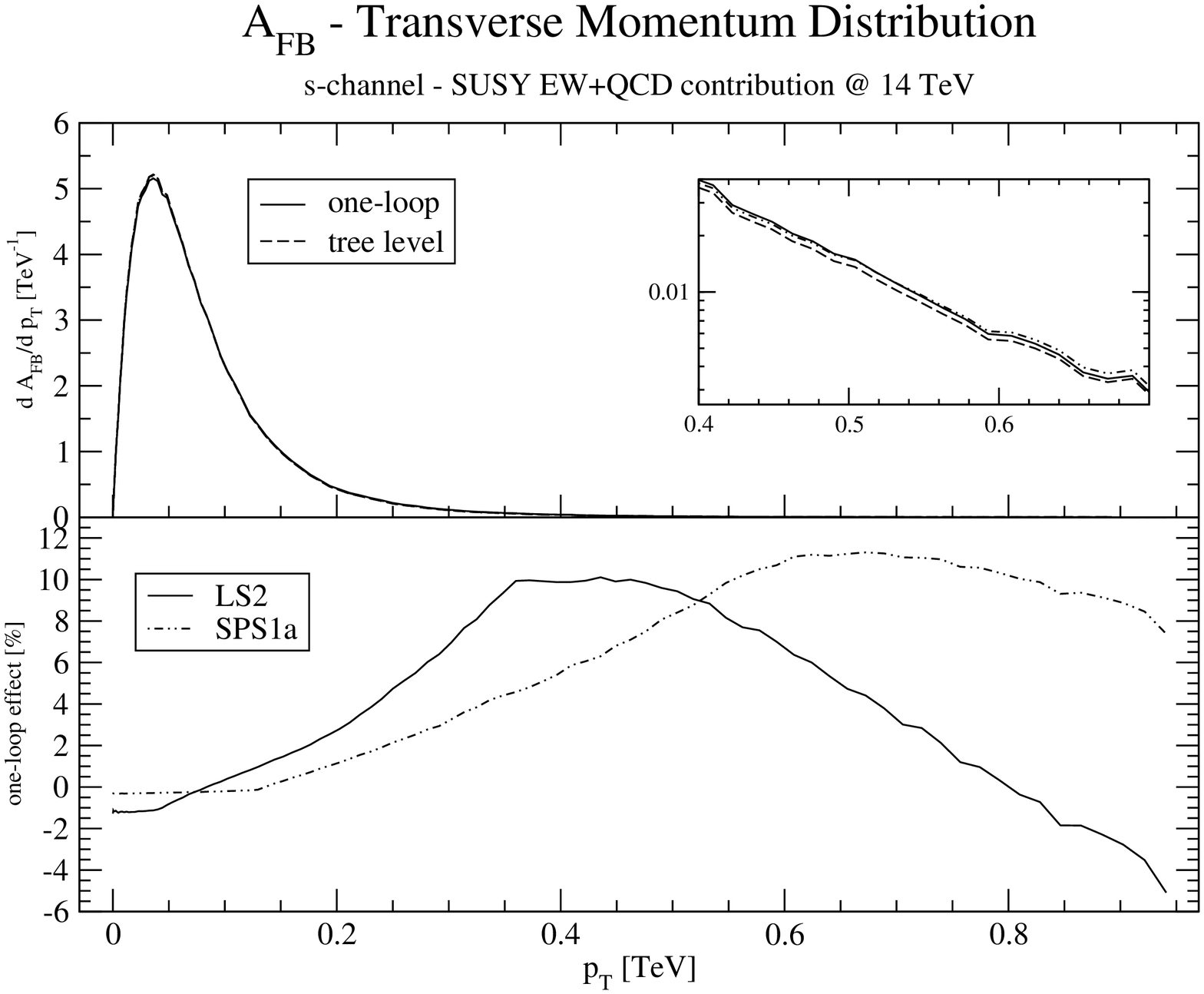,
width=0.48\textwidth, angle=0}\\[35pt]
\epsfig{file=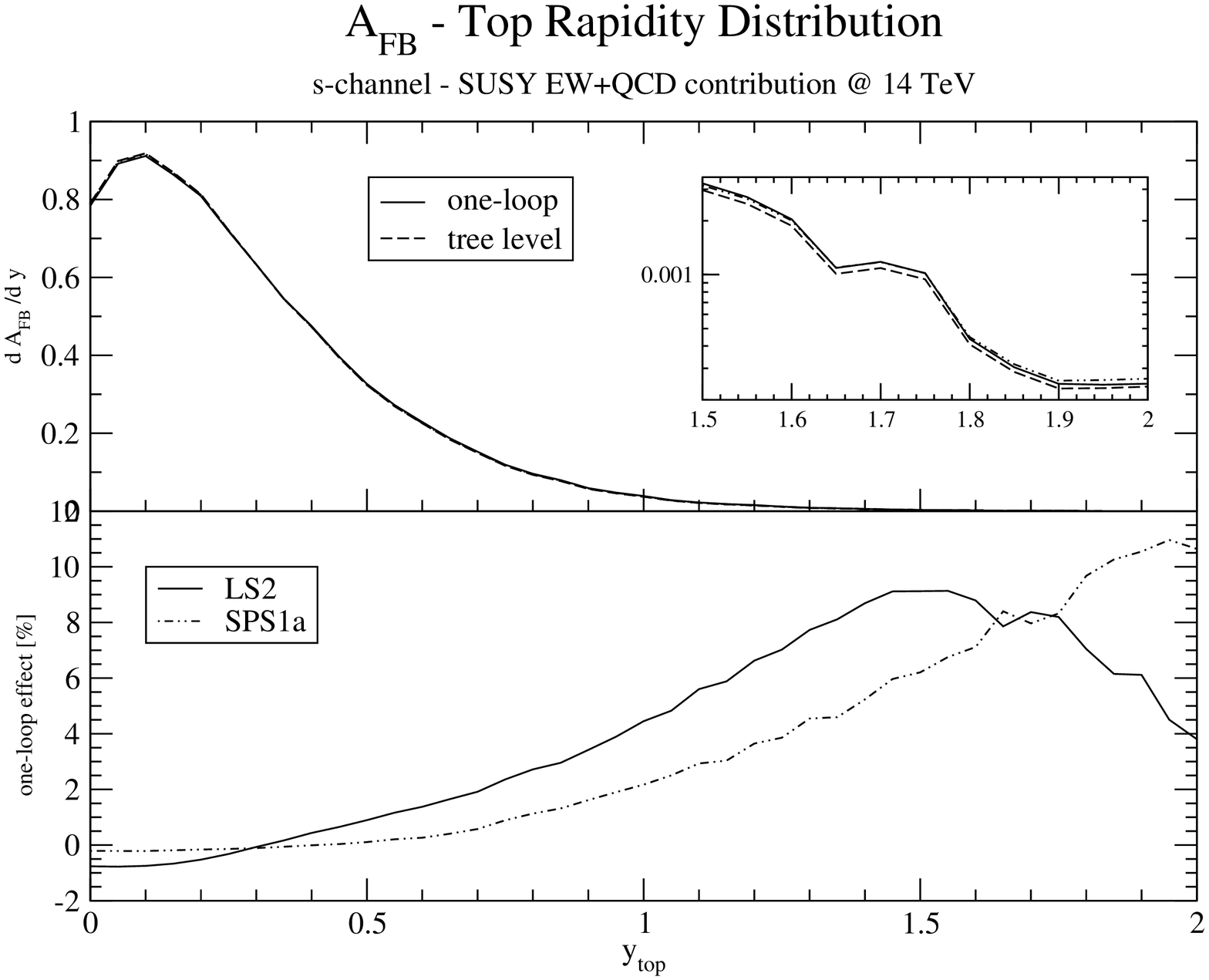,
width=0.48\textwidth, angle=0}\hfill
\epsfig{file=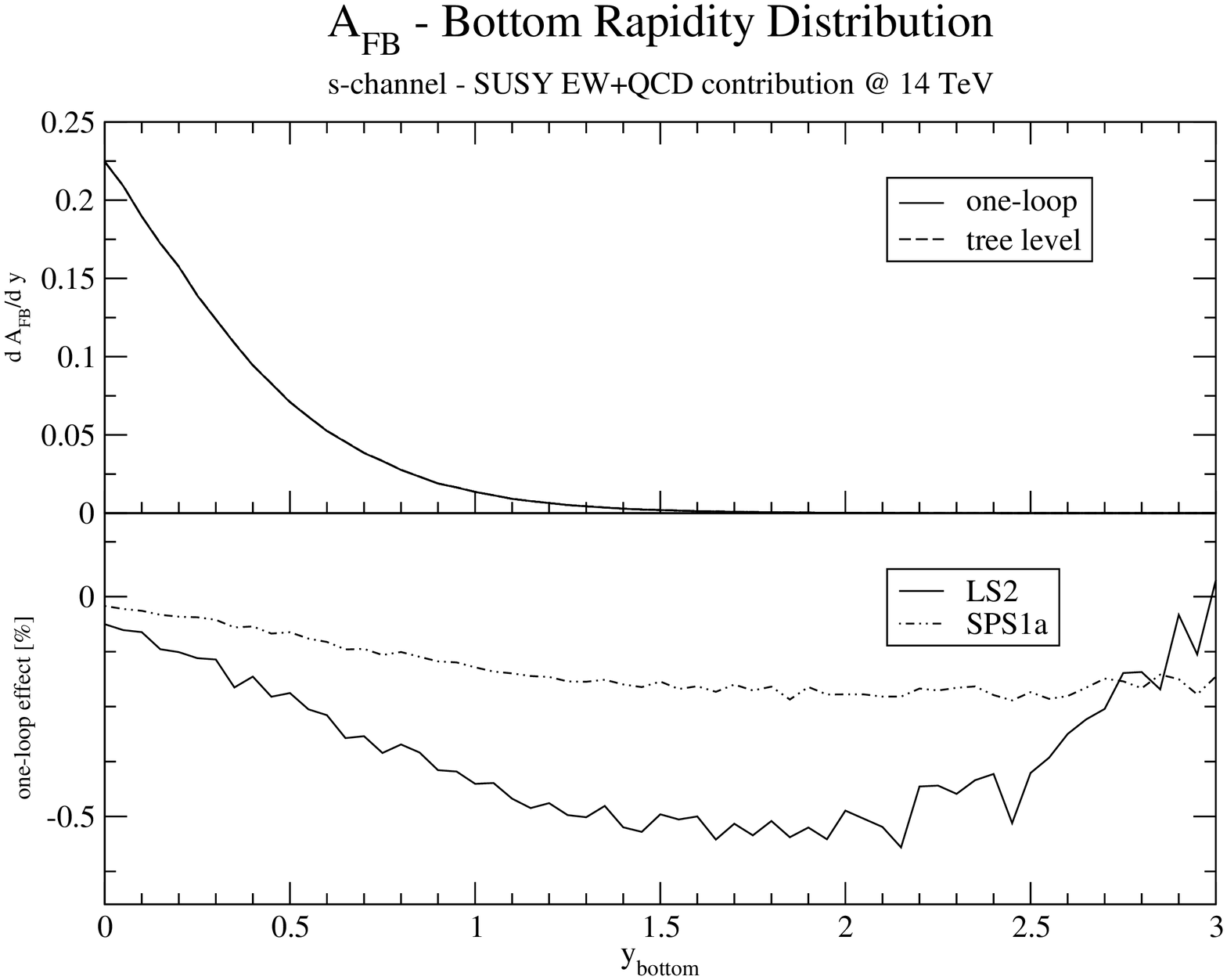,
width=0.48\textwidth, angle=0}\\[35pt]
\caption{Differential distributions of the Forward-Backward asymmetry in $s$-channel at 14 TeV.}
\label{fig:AFB-s-channel-14TeV}
\end{figure}

The corrections to differental quantities in the Forward-Backward asymmetry are qualitatively
and quantitatively very similar to those for the Left-Right asymmetry, as it is possible to see in
Fig.~\ref{fig:AFB-s-channel-14TeV}, thus the same comments as above apply in this case.

Considering all the differential distributions for the s-channel process, we can notice that they
do not differ too much from each other between the two benchmarks and, as a general feature, they reach their maximum value in
the invariant mass distribution for $M_{\rm inv}\gtrsim 700$~GeV, in the transverse momentum
distribution for $p_T\gtrsim 300$~GeV and in the top rapidity for $y_t \gtrsim 1.5$.\\

\begin{figure}[t]
\centering
\epsfig{
file=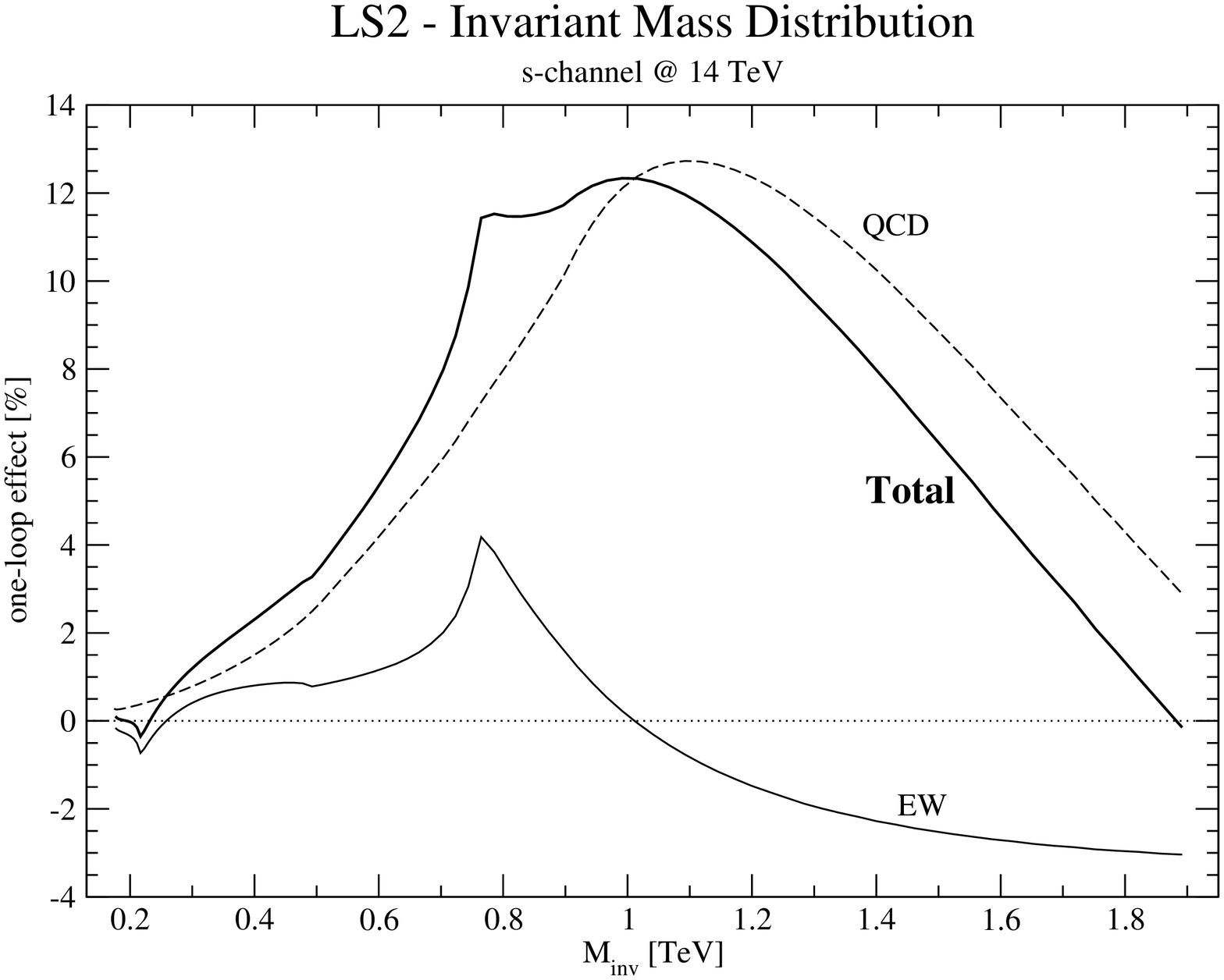,
width=0.48\textwidth, angle=0}\hfill
\epsfig{
file=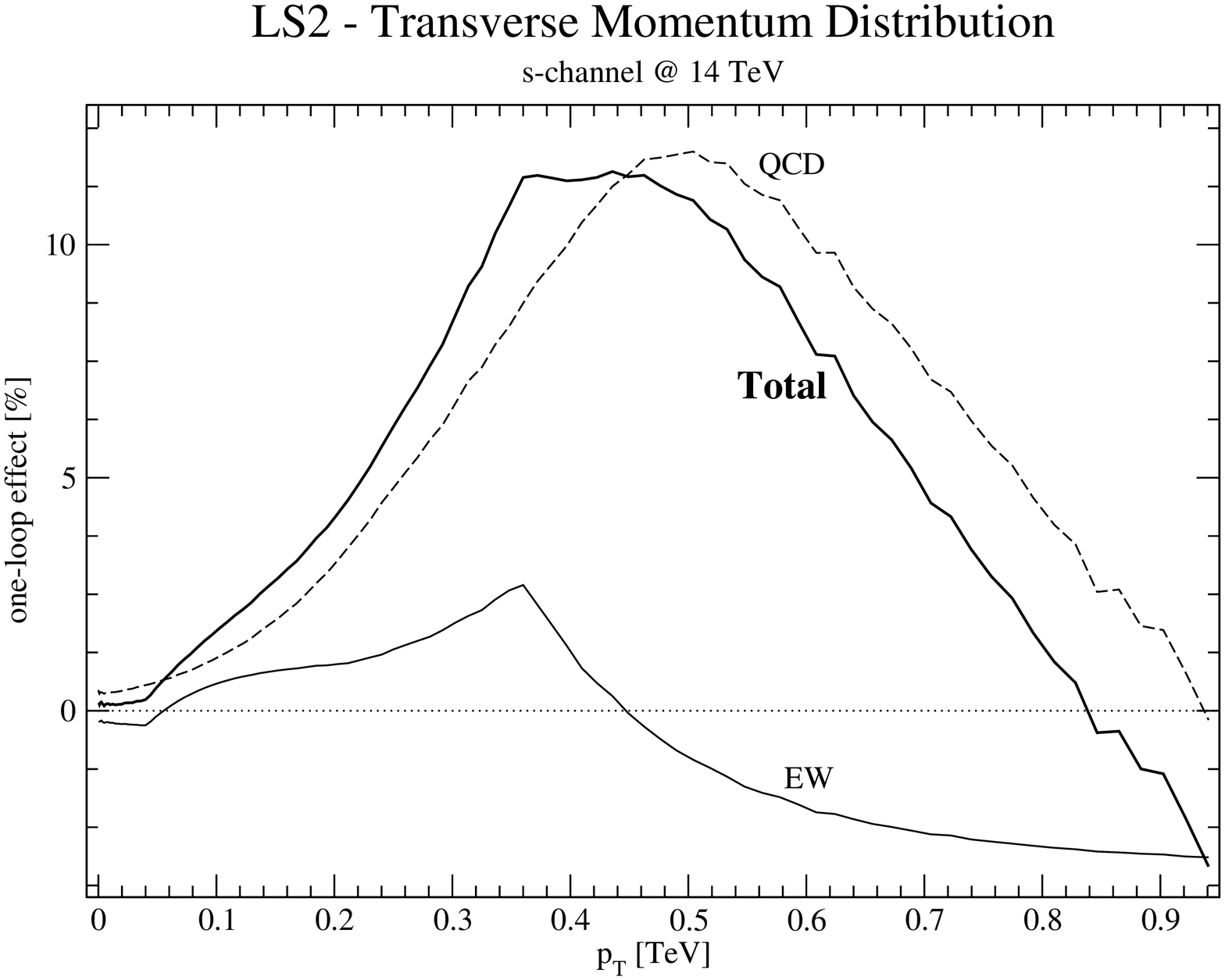,
width=0.48\textwidth, angle=0}\\[35pt]
\epsfig{
file=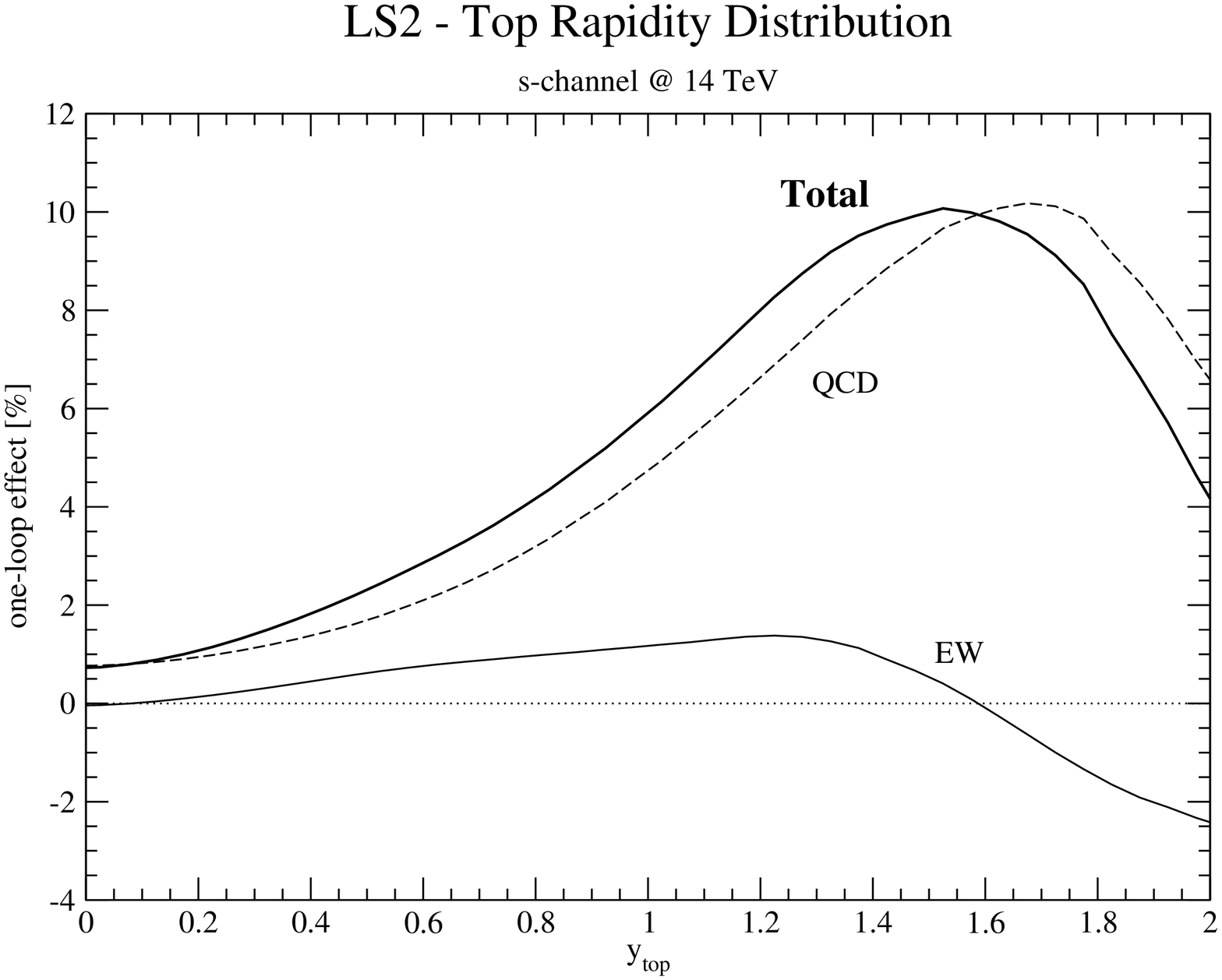,
width=0.48\textwidth, angle=0}\hfill
\epsfig{
file=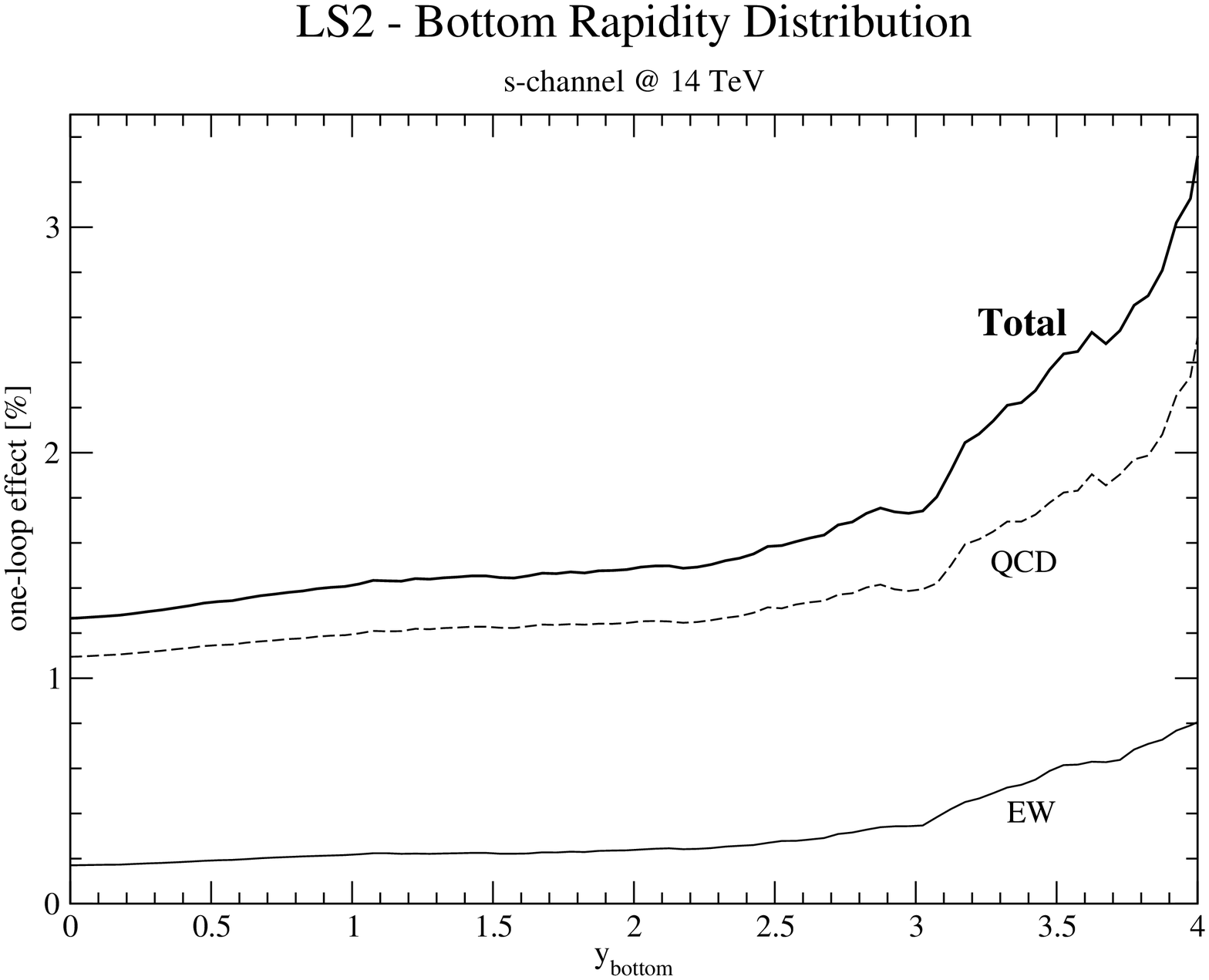,
width=0.48\textwidth, angle=0}\\[35pt]
\caption{SUSY EW and QCD corrections to differential distributions of the cross section in $s$-channel at 14 TeV for LS2.}
\label{fig:EWQCD-s-channel-LS2}
\end{figure}

\begin{figure}[t]
\centering
\epsfig{
file=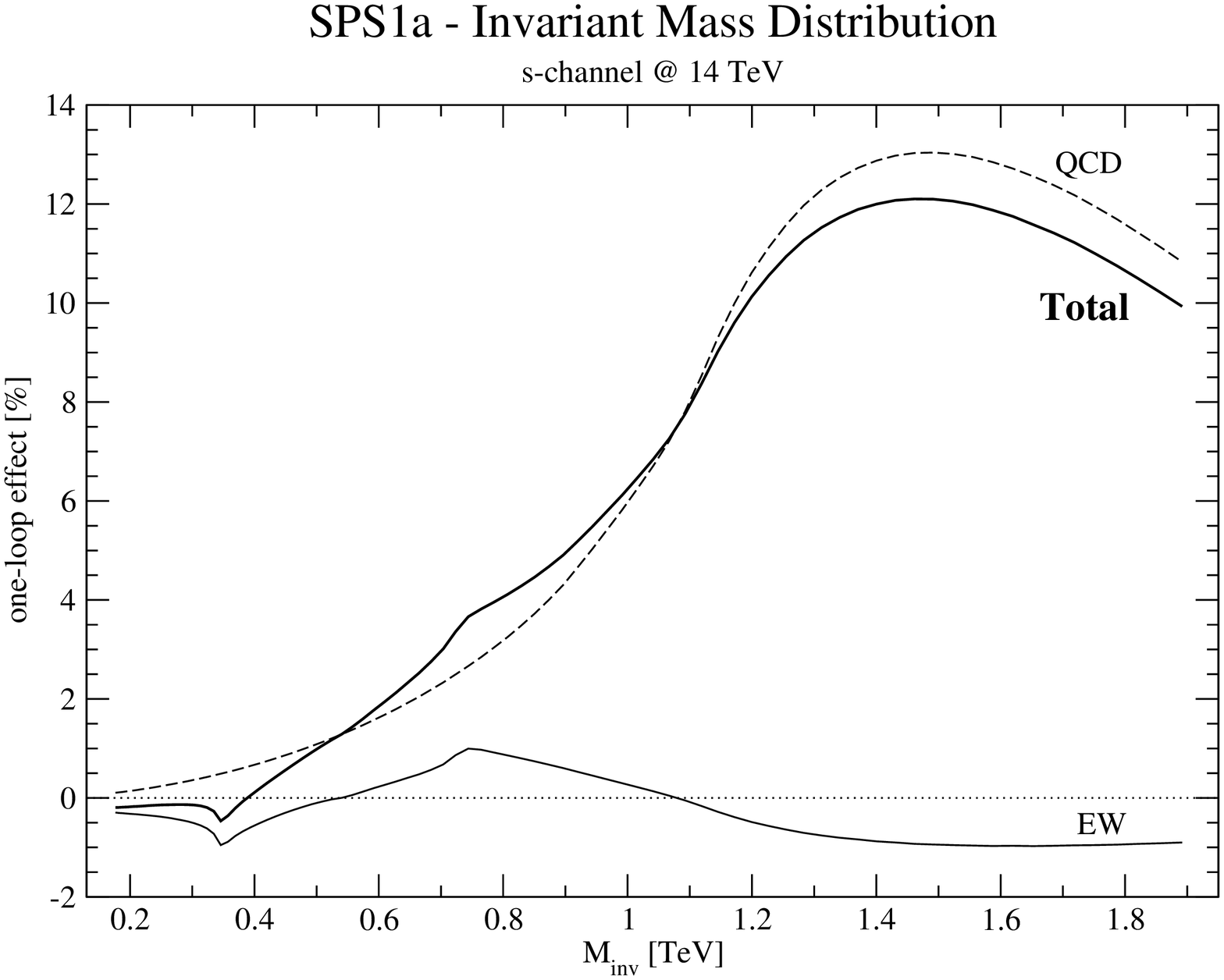,
width=0.48\textwidth, angle=0}\hfill
\epsfig{
file=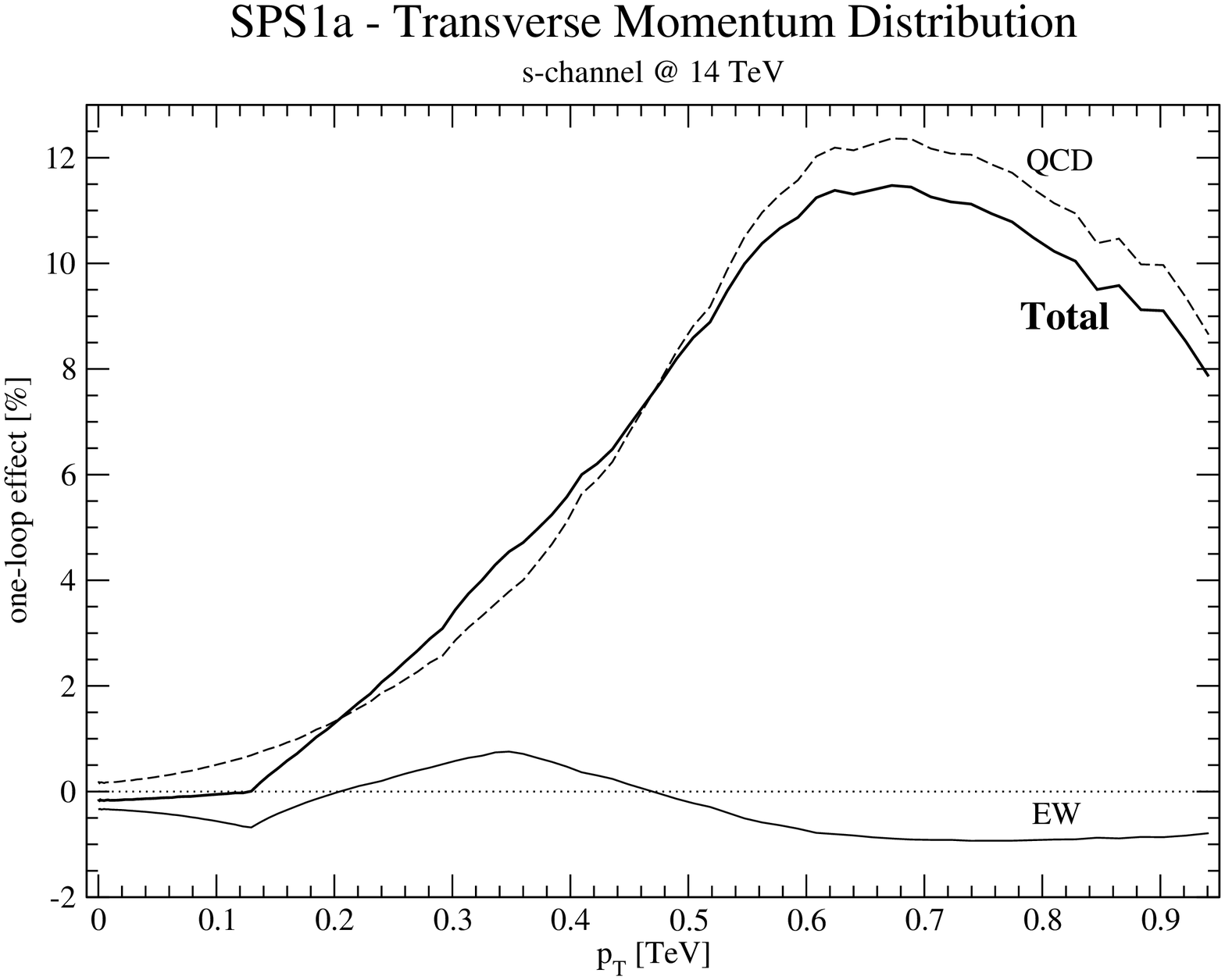,
width=0.48\textwidth, angle=0}\\[35pt]
\epsfig{
file=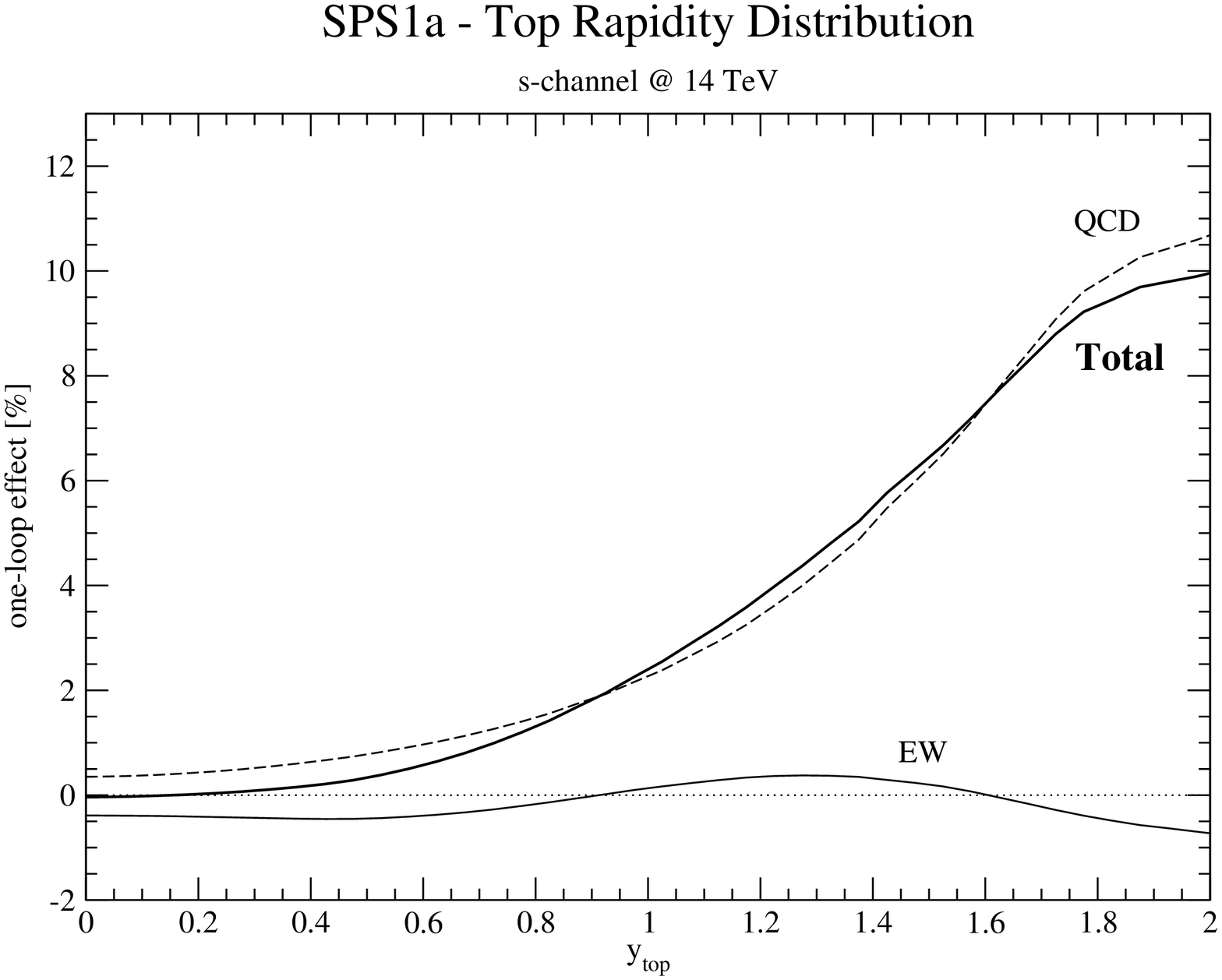,
width=0.48\textwidth, angle=0}\hfill
\epsfig{
file=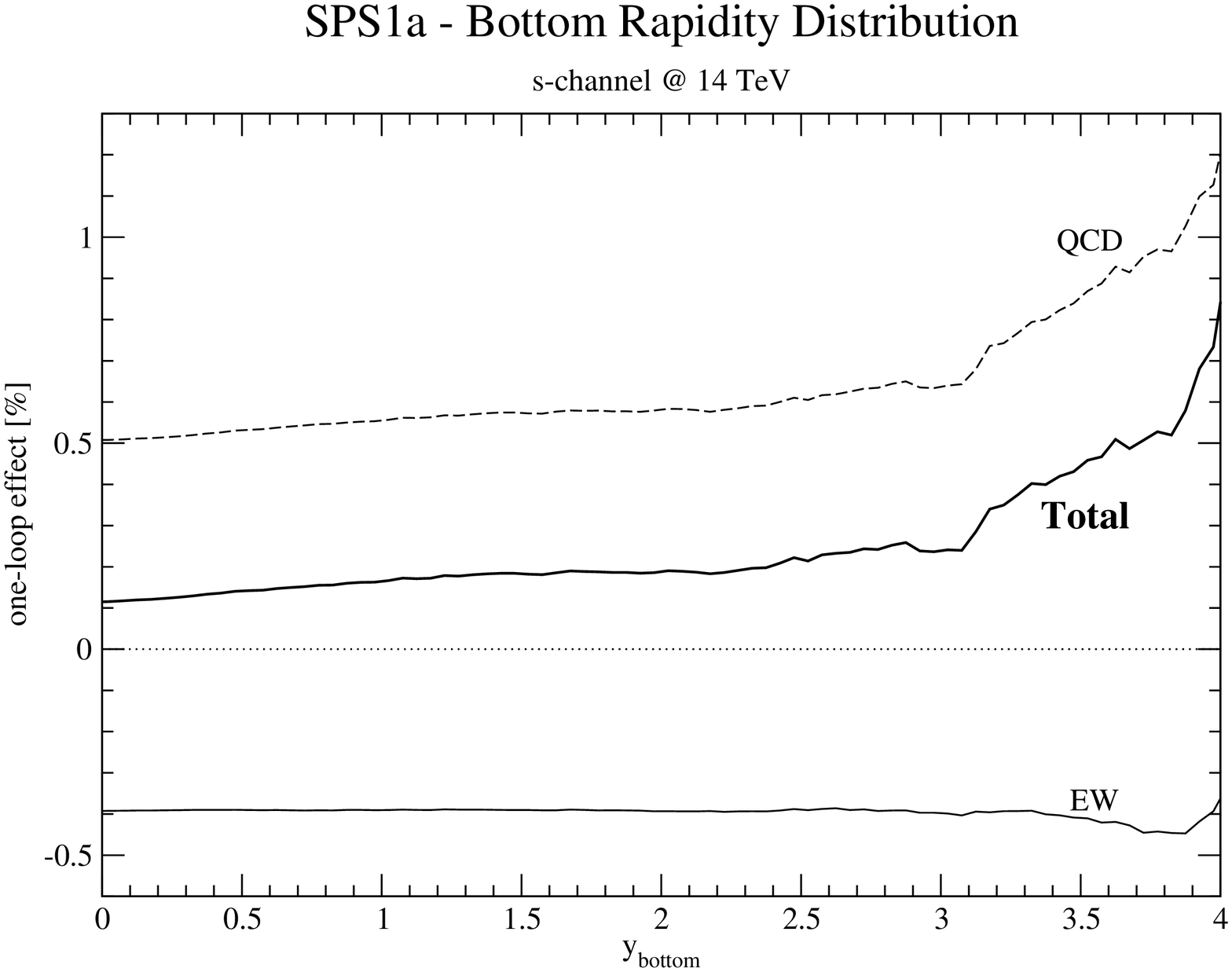,
width=0.48\textwidth, angle=0}\\[35pt]
\caption{SUSY EW and QCD corrections to differential distributions of the cross section in $s$-channel at 14 TeV for SPS1a.}
\label{fig:EWQCD-s-channel-SPS1}
\end{figure}

To better understand the origin of the corrections and why there are peaks for certain
values of $M_{\rm inv}$, $p_T$ and $y_t$ we have computed (here, limitedly to the s-channel) the
SUSY EW and QCD corrections
for the cross section separately and plotted the related differential quantities in
Figs.~\ref{fig:EWQCD-s-channel-LS2} and~\ref{fig:EWQCD-s-channel-SPS1} for the two benchmark scenarios. It is possible to see that
the corrections are dominated by
the SUSY QCD contribution, while the EW counterpart 
provides small
contributions to the total corrections. In both benchmark scenarios, however, the EW contribution shows
peculiar features in the invariant mass distribution, given by peaks and troughs in the correction,
and a closer inspection of these threshold effects reveals that they are situated in correspondence
of $M_{\chi_i^\pm}^2$. Therefore charginos could play an interesting role in the determination of
SUSY effects in s-channel single top production processes. Going back to the dominant QCD
contribution, the only parameter which enter the QCD correction alone is
the gluino mass ($m_{\tilde g} = 607$~GeV in SPS1a and $m_{\tilde g} = 392$~GeV in
LS2), thus we can argue that the one-loop corrections are sensitive mostly to this parameter.
Smaller gluino masses shift the SUSY QCD peak in the corrections towards regions where the 
differential distributions have higher values and therefore the integrated quantities will
be affected more by NLO corrections. This means that observing higher pure SUSY one-loop
corrections would point towards scenarios with lighter gluino and viceversa.\\

\begin{figure}[ht]
\centering
\epsfig{file=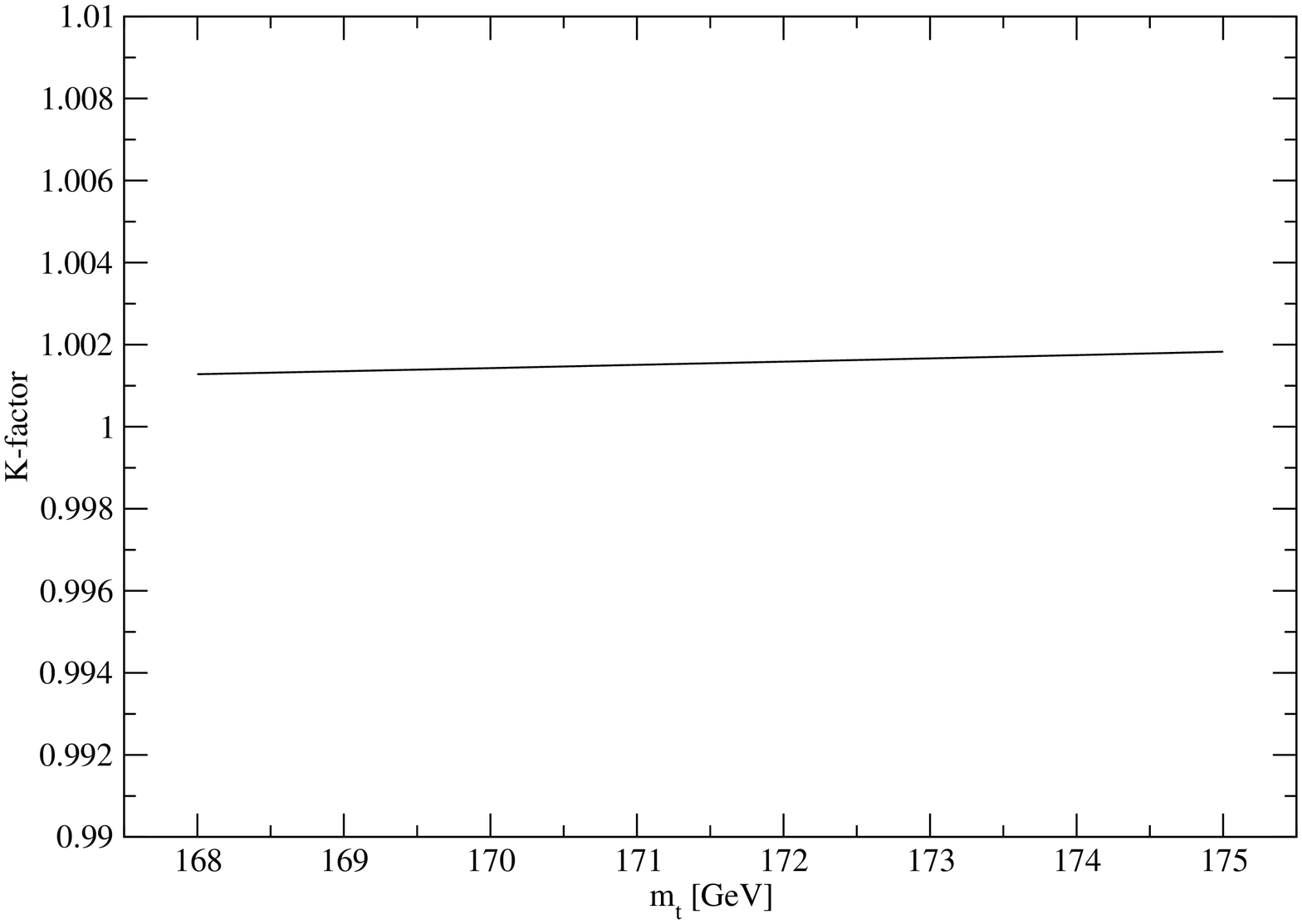, width=0.8\textwidth, angle=0}\\[50pt]
\caption{K-factor dependence on the top mass in s-channel at 14 TeV for SPS1a.}
\label{fig:topmass}
\end{figure}

To conclude this section, we are aware that recent measurements have slightly changed the value of
the top mass\cite{topmass}. Nevertheless, we have verified that any change (within a reasonable
range) in this value has practically no effect on one-loop corrections, as it is possible to see in
Fig.~\ref{fig:topmass}.




\section{Conclusions}
\label{Sec:Conclusions}
In summary, QCD and EW corrections through one-loop level originating from SUSY and affecting the cross section for `$t$~+~jet' final states at the LHC are always small at inclusive level, at both energies of 7 and 14 TeV. However, they appear
sizable (up to ten percent) and detectable (after ${\cal O}(100)$ fb$^{-1}$ of luminosity)
at differential level, if the collider energy is 14 TeV, while for 7 TeV they will presumably not be
accessible. Such large corrections also appear in the
Left-Right and Forward-Backward (when defineable) asymmetries,
again, at the differential but not inclusive level. The bulk of the
higher order effects comes from QCD, with a much smaller, yet 
very sensitive to SUSY parameters, EW component. The corrections
discussed here are typically larger for the s-channel mode than for the t-channel one, which
renders them more difficult to access as the former is subleading with respect to the latter at the
LHC. Altogether though, our results point to the relevance of SUSY one-loop effects onto `$t$~+~jet'
final states for the LHC running at design energy and luminosity. In presence of high statistic
samples, some typical SUSY parameter dependencies could possibly be extracted,
in either mSUGRA or the MSSM.


\section*{Acknowledgements} 
Both GM and LP acknowledge financial support from the Royal Society (London, UK)
in the form of a travel grant to visit Southampton. SM is financially supported in part by 
the scheme `Visiting Professor - Azione D - Atto Integrativo tra la 
Regione Piemonte e gli Atenei Piemontesi'. 
GM and LP would like to thank Matteo Beccaria, Claudio Verzegnassi, and Fernand Renard for
valuable suggestions and discussions during earlier work on single top production.



\hfill

\hfill

\hfill

\hfill

\hfill

\hfill

\hfill

\hfill

\hfill

\hfill


\begin{thebibliography}{99}



\bibitem{QCD-SM-tt}

  W.~Beenakker, W.~L.~van Neerven, R.~Meng, G.~A.~Schuler and J.~Smith,
  Nucl.\ Phys.\  B {\bf 351} (1991) 507;

  W.~Bernreuther, A.~Brandenburg, Z.~G.~Si and P.~Uwer,
  Int.\ J.\ Mod.\ Phys.\  A {\bf 18} (2003) 1357; 

  N.~Kidonakis and R.~Vogt,
  Phys.\ Rev.\  D {\bf 78} (2008) 074005;

  M.~Cacciari, S.~Frixione, M.~L.~Mangano, P.~Nason and G.~Ridolfi,
  JHEP {\bf 0809} (2008) 127;

  M.~Czakon and A.~Mitov,
  Nucl.\ Phys.\  B {\bf 824} (2010) 111;

  K.~Melnikov and M.~Schulze,
  JHEP {\bf 0908} (2009) 049;

  A.~Bredenstein, A.~Denner, S.~Dittmaier and S.~Pozzorini,
  JHEP {\bf 1003} (2010) 021.

\bibitem{EW-SM-tt}

  S.~Moretti, M.~R.~Nolten and D.~A.~Ross,
  Phys.\ Lett.\  B {\bf 639} (2006) 513
  [Erratum-ibid.\  B {\bf 660} (2008) 607];

  J.~H.~Kuhn, A.~Scharf and P.~Uwer,
  Eur.\ Phys.\ J.\  C {\bf 51} (2007) 37;

  W.~Hollik and M.~Kollar,
  Phys.\ Rev.\  D {\bf 77} (2008) 014008;

  W.~Bernreuther, M.~Fucker and Z.~G.~Si,
  Nuovo Cim.\  {\bf 123B} (2008) 1036.

\bibitem{QCD-SM-t} 

  B.~W.~Harris, E.~Laenen, L.~Phaf, Z.~Sullivan and S.~Weinzierl,
  Phys.\ Rev.\  D {\bf 66} (2002) 054024;


  S.~Frixione, E.~Laenen, P.~Motylinski and B.~R.~Webber,
  JHEP {\bf 0603} (2006) 092;

  N.~Kidonakis,
  Phys.\ Rev.\  D {\bf 75} (2007) 071501;

  S.~Frixione, E.~Laenen, P.~Motylinski, B.~R.~Webber and C.~D.~White,
  JHEP {\bf 0807} (2008) 029;

  J.~M.~Campbell, R.~Frederix, F.~Maltoni and F.~Tramontano,
  Phys.\ Rev.\ Lett.\  {\bf 102} (2009) 182003;

  S.~Heim, Q.~H.~Cao, R.~Schwienhorst and C.~P.~Yuan,
  Phys.\ Rev.\  D {\bf 81} (2010) 034005.

\bibitem{EW-SM-t}

  M.~Beccaria, C.~M.~Carloni Calame, G.~Macorini, E.~Mirabella, F.~Piccinini,
  F.~M.~Renard and C.~Verzegnassi,
  Phys.\ Rev.\  D {\bf 77} (2008) 113018;

  M.~Beccaria, G.~Macorini, F.~M.~Renard and C.~Verzegnassi,
  Phys.\ Rev.\  D {\bf 74} (2006) 013008;

  M.~Beccaria, C.~M.~Carloni Calame, G.~Macorini, G.~Montagna, F.~Piccinini,
  F.~M.~Renard and C.~Verzegnassi,
  Eur.\ Phys.\ J.\  C {\bf 53} (2008) 257.

\bibitem{QCD-MSSM-tt}
  D.~A.~Ross and M.~Wiebusch,
  JHEP {\bf 0711} (2007) 041.


\bibitem{EW-MSSM-tt}
  W.~Bernreuther, M.~Fuecker and Z.~G.~Si,
  Phys.\ Rev.\  D {\bf 74} (2006) 113005.


\bibitem{EW-MSSM-t}

  M.~Beccaria, F.~M.~Renard and C.~Verzegnassi,
  Phys.\ Rev.\  D {\bf 71} (2005) 033005;

  C.~S.~Li, R.~J.~Oakes and J.~M.~Yang,
  Phys.\ Rev.\  D {\bf 55} (1997) 5780.


\bibitem{Pumplin:2002vw}
  J.~Pumplin, D.~R.~Stump, J.~Huston, H.~L.~Lai, P.~M.~Nadolsky and W.~K.~Tung,
  JHEP {\bf 0207} 012 (2002) 012.


\bibitem{Djouadi:2002ze}
  A.~Djouadi, J.~L.~Kneur and G.~Moultaka,
  Comput.\ Phys.\ Commun.\  {\bf 176} (2007) 426.


\bibitem{Barger:1986hd}
  V.~D.~Barger, N.~G.~Deshpande, J.~L.~Rosner and K.~Whisnant,
  Phys.\ Rev.\  D {\bf 35} (1987) 2893.



\bibitem{Brein:2004kh}
  O.~Brein,
  Comput.\ Phys.\ Commun.\  {\bf 170} (2005) 42.


\bibitem{Allanach:2002nj}
  B.~C.~Allanach {\it et al.},
Eur.\ Phys.\ J.\  C {\bf 25} (2002) 113.

\bibitem{Beccaria:2006ir}
  M.~Beccaria, G.~Macorini, F.~M.~Renard and C.~Verzegnassi,
  Phys.\ Rev.\  D {\bf 74} (2006) 013008.

\bibitem{topmass}
K.~Nakamura  et al. (Particle Data Group), Journal of Physics G {\bf 37}, 075021 (2010)


\end{thebibliography}
\end{document}